\documentclass[a4paper,11pt]{article}

\usepackage{latexsym}
\usepackage{amsfonts} 
\usepackage[mathscr]{euscript}
\usepackage{amsmath,mathrsfs,bm,amssymb,color}

\title{\textsf{
Existence  of charge-density waves in  two-dimensional ionic Hubbard model
}}
\date{\empty}
\date{\empty}
\author{
Tadahiro Miyao\\ 
 {\it Department of Mathematics,}
{\it Hokkaido University,}\\
{\it Sapporo 060-0810, Japan}\\
E-mail:
 miyao@math.sci.hokudai.ac.jp
}

\newcommand{\one}{{\mathchoice {\rm 1\mskip-4mu l} {\rm 1\mskip-4mu l}
{\rm 1\mskip-4.5mu l} {\rm 1\mskip-5mu l}}}
\newcommand{\h}{\mathfrak{h}}

\newcommand{\Fock}{\mathfrak{F}}

\newcommand{\ran}{\mathrm{ran}}

\newcommand{\la}{\langle}
\newcommand{\ra}{\rangle}
\newcommand{\Tr}{\mathrm{Tr}}

\newcommand{\BbbR}{\mathbb{R}}
\newcommand{\BbbN}{\mathbb{N}}
\newcommand{\BbbZ}{\mathbb{Z}}

\newcommand{\BbbC}{\mathbb{C}}
\newcommand{\vepsilon}{\varepsilon}
\newcommand{\vphi}{\varphi}

\newcommand{\no}{\nonumber \\}

\newcommand{\mb}{\mathbf}
\newcommand{\mr}{\mathrm}

\newcommand{\bi}{{\boldsymbol i}}
\newcommand{\bj}{{\boldsymbol j}}
\newcommand{\bfm}{{\boldsymbol m}}
\newcommand{\bn}{{\boldsymbol n}}

\begin{document}

\newtheorem{define}{Definition}[section]
\newtheorem{Thm}[define]{Theorem}
\newtheorem{Prop}[define]{Proposition}
\newtheorem{lemm}[define]{Lemma}
\newtheorem{rem}[define]{Remark}
\newtheorem{assum}{Condition}
\newtheorem{example}{Example}
\newtheorem{coro}[define]{Corollary}

\maketitle
\begin{abstract}
We rigorously investigated the charge-charge correlation function of the   ionic Hubbard
 model in two dimensions  by reflection positivity.
We prove the existence of charge-density waves 
 for large staggered potential $\Delta$ (i.e., $\frac{\Delta}{2}+2V>U$) at low temperatures, 
where $U$ and $V$ are  the on-site  and nearest-neighbor Coulomb
 repulsions, respectively.
The results are consistent  with  previous  numerical simulation results. 
We argue that the absence of charge-density waves  for $\Delta=0$ and
 $U$ are 
 large enough (i.e., $U>\frac{\Delta}{2}$+2V).
\end{abstract}
\tableofcontents

\section{Introduction}
\setcounter{equation}{0}
The ionic Hubbard model was originally suggested   to describe the charge-transfer
organic salts \cite{HT,NT},  although it was subsequently used  to analyze 
  ferroelectric perovskites \cite{EIT, FGN, RS}.
In these studies, the half-filled one-dimensional model was used to  
 understand   quantum phase transitions from 
 band insulators to  Mott insulators. 
The ionic Hubbard model comprises  the usual Hubbard model with on-site 
Coulomb repulsion $U$ supplemented by an alternating one-particle potential 
of magnitude $\Delta$.
Surprisingly,  this model is reported to have  two qunatum  critical points
as  $U$ is varied with  fixed $\Delta$ \cite{BA,KSJ, MMN,PBH}.
For $U<U_1$, the system is a band insulator. In the intermediate regime $U_1<U<U_2$, 
the system has a bond-order characterized by  the ground-state expectation value  
of the staggered kinetic energy per bond. 
For $U>U_2$, the system is a Mott insulator.

Recently, the  two-dimensional ionic Hubbard model has received significant attention both theoretically and 
experimentally.
However , the  phase diagram of this  model  remains a mystery.
Conversely,  recently developed  experimental techniques 
make it possible to implement the ionic Hubbard model in an optical honeycomb lattice.
As theoretical studies have  suggested,
for large $\Delta$, a charge-density wave appears. For large $U$, the
charge-density wave is strongly suppressed \cite{MDU}. 
Whether  bond-order exists  in a  two-dimensional system, however,
remains unclear. 

Many  theoretical studies  are based on numerical simulations that
 clarify
the  properties of the ionic Hubbard model in one  and two dimensions; 
however, limited exact results are available.
The aim of  the present work was to  study the extended ionic Hubbard
model in a  two-dimensional square lattice rigorously using {\it reflection positivity}. 
The results of this approach prove the existence of a staggered long-range charge order for
$\frac{\Delta}{2}+2V>U$, where $V$ is the nearest-neighbor Coulomb repulsion.
This finding  justifies  a part of the phase diagram suggested  by 
numerical simulations \cite{BPH,DK, Murakami,PBH}.
Although we do   not address  the bond-order in this paper,  we discuss some rigorous results for
$U>\frac{\Delta}{2}+2V$. The work constitutes  the  first step in 
the rigorous study  of the two-dimensional extended 
 ionic Hubbard model. 

A similar reflection-positive-based  model for   three and more dimensions
was previously discussed \cite{FILS2}.
However,  the present work is restricted to   two-dimensional models.

The proposed method works well only for a sufficiently
small 
hopping amplitude $t$.
We are  aware of no rigorous results when $t$ is large (i.e., $U \approx t$).

Reflection positivity originates from axiomatic quantum field theory \cite{OS}.
Glimm, Jaffe, and Spencer found that  reflection positivity can be
applied   to the rigorous 
study of  phase transitions \cite{GJS,GlimmJaffe}.
In the 1970s, Dyson, Fr\"ohlich, Israel, Lieb, Spencer,  and  Simon 
established the  foundation of the methods of reflection positivity in  statistical physics \cite{DLS,FL,FILS,FILS2,FSS}.
Reflection positivity has been successfully applied to numerous  models
and is  regarded as a crucial analysis method in condensed-matter
physics \cite{BK,BiCS,vES,SY}.

In the present work, we adapted  the  method by  Fr\"ohlich and Lieb \cite{FL} to 
the ionic Hubbard model.
In \cite{FL}, Fr\"ohlich and Lieb proved  the existence of long-range
order  in the two-dimensional
 models,  including the anisotropic Heisenberg model.  
Their proof contains  the following three parts:
\begin{itemize}
\item[(i)] the Peierls argument;
\item[(ii)] the chessboard estimate;
\item[(iii)] the principle of exponential localization.
\end{itemize} 
Reflection positivity is a basic input of the chessboard estimate. 
To use the chessboard estimate,   the model was required to be 
defined on  a square lattice with sides of length $4M$ in \cite{FL}.
In contrast, our  reflection positivity arguments work well in a square lattice with sides of length $4M+2$.
This difference requires several  extensions of  the Fr\"ohlich--Lieb method.
Our methodological achievement is that we actually completed these extensions.

Note that our result could be proven by the quantum Pirogov--Sinai
theory \cite{BKU, DFF}.  However, as far as we know,  there is no proof  based on
reflection positivity.

The organization of the paper is as follows:
 In Section \ref{Result}, we present the definition of the  ionic
 Hubbard model and state the main results.

In Section \ref{Stra}, we describe  the  strategy of the main theorem (Theorem \ref{MainResult}).

Sections \ref{PeiArg}--\ref{Proof(B)} are devoted to the proof of Theorem \ref{MainResult}.

In Appendix \ref{DLS}, we present the proof of  an extension of the
Dyson--Lieb--Simon (DLS) inequality.

\begin{flushleft}
{\bf Acknowledgements:}
This work was supported by KAKENHI(16H03942).
\end{flushleft}

\section{Results}\label{Result}
\setcounter{equation}{0}

Let $\Lambda=[-L, L)^2\cap \BbbZ^2$ with $L\in \BbbN$.
The extended ionic Hubbard model is 
\begin{align}
H=&(-t)\sum_{\la \bi, \bj\ra}\sum_{\sigma=\uparrow, \downarrow}\big(
c_{\bi\sigma}^* c_{\bj\sigma}
+c_{\bj\sigma}^* c_{\bi\sigma}
\big)\no
&+U\sum_{\bj\in \Lambda} (n_{\bj}-\one)^2+V\sum_{\la \bi, \bj\ra}
 (n_{\bi}-\one)
 (n_{\bj}-\one)
+\frac{\Delta}{2}\sum_{\bj\in \Lambda} (-1)^{|\bj|}(n_{\bj}-\one),
\end{align} 
where $
\sum_{\la \bi, \bj\ra}
$
 means a sum over nearest neighbors;
$t$ is the hopping amplitude between nearest-neighbor;
$U$ and $V$ define the on-site and nearest-neighbor Coulomb interactions, respectively;
$\Delta$ denotes an alternating chemical potential; and
$c_{\bj\sigma} (c_{\bj\sigma}^*)$ is the standard fermion annihilation(creation) operator on  site $\bj=(j_1, j_2)\in \Lambda$  with spin
$\sigma$.
The number operator $n_{\bj}$ is defined by 
$n_{\bj}=n_{\bj\uparrow}+n_{\bj\downarrow}$ with $n_{\bj\sigma}=c_{\bj\sigma}^*c_{\bj\sigma}$.
$H$ acts on the fermion Fock space $\mathfrak{H}=\Fock(\ell^2(\Lambda) \oplus
\ell^2(\Lambda))$, where $
\Fock(\mathfrak{h})=\bigoplus_{n\ge 0}\wedge^n \mathfrak{h}
$. Here $\wedge^n \mathfrak{h}$ is the $n$-fold antisymmetric tensor
product of $\mathfrak{h}$ with $\wedge^0\mathfrak{h}=\BbbC$.
We impose a periodic boundary condition, so $L\equiv -L$.

In the present paper, we assume the following:
\begin{itemize}
\item $\Delta \ge 0,\ \ V\ge 0,\ \ U\in \BbbR$.
\item $t>0$.
\item $L$ is odd.
\end{itemize}

The thermal average is defined by 
\begin{align}
\la A \ra_{\beta, \Lambda, H}=\Tr_{\mathfrak{H}}\big[A\, e^{-\beta H}\big]\Big/
Z_{\beta, \Lambda, H},\ \ \ 
Z_{\beta, \Lambda, H}=\Tr_{\mathfrak{H}}\big[ e^{-\beta H}\big].
\end{align} 
Note  that  if $\Delta=0$, then the system is half-filled.
Let $q_{\bj}=n_{\bj}-\one$. The charge-charge  correlation function is given
by
\begin{align}
\la q_{{\bf o}} q_{\bj}\ra_{\beta, H}=\lim_{L\to \infty}
\la q_{{\bf o}} q_{\bj}\ra_{\beta, \Lambda, H}.
\end{align} 
The main result in this paper is the following:
\begin{Thm}\label{MainResult}
If $\displaystyle
2V-U+\frac{\Delta}{2}>0
$, then, for  sufficiently large $\beta$ and small $t$, 
\begin{align}
\liminf_{|\bj|\to \infty} (-1)^{|\bj|} \la q_{{\bf o}} q_{\bj}\ra_{\beta,
 H}>0, \label{LROClaim}
\end{align}
where $|\bj|=|j_1|+|j_2|$ for each $\bj=(j_1, j_2)\in \Lambda$. 
Thus, there exists a long-range charge order.
\end{Thm}

\begin{rem}
{\rm 

If $V=0$, then our result are consistent with  the phase diagrams  obtained using    numerical
 simulations \cite{BPH,DK,PBH}. 
On the other hand, if $\Delta=0$, then Theorem \ref{MainResult} agrees
 with  results predicted  by numerical simulations, see,  e.g., \cite{Murakami}.
$\diamondsuit$
}
\end{rem} 

\begin{rem}
{\rm 
In three or more dimensions, we can prove the existence  of long-range
 charge order  by the method established in \cite{DLS,FILS,FILS2,T. Miyao5}. $\diamondsuit$ 
}
\end{rem} 

\begin{rem}
{\rm 
Assume $\Delta=0$.  If $2V-U<0$,  then we already know the following:
\begin{itemize}
\item[{\rm (i)}] For all $\beta$ and $t$, 
\begin{align}
\lim_{|\bj|\to \infty} (-1)^{|\bj|} \la q_{{\bf o}} q_{\bj}\ra_{\beta, H}=0.
\end{align} 
Thus, there is no long-range charge order \cite{KuboKishi,Miyao3,Miyao4}.
\item[{\rm (ii)}] If $L$ finite, the  ground state of $H$  is
	     unique and 
	     antiferromagnetic \cite{Lieb,Miyao2,Miyao3}. $\diamondsuit$
\end{itemize} 
}
\end{rem}

\section{Outline of proof of Theorem \ref{MainResult}} \label{Stra}
\setcounter{equation}{0}

\subsection{Preliminaries}

Let 
\begin{align}
v_{\bj\sigma}=\Bigg[
\prod_{\bi\neq \bj} (-1)^{n_{\bi\sigma}} 
\Bigg](c_{\bj\sigma}^*+c_{\bj\sigma}).
\end{align} 
It is not hard to check that 
\begin{align}
v_{\bi\sigma} c_{\bj\sigma'} v_{\bi\sigma}^{-1} =\begin{cases}
c_{\bi\sigma}^ * & \mbox{if $(\bi, \sigma)=(\bj, \sigma')$}\\
c_{\bj\sigma'} & \mbox{if  $(\bi, \sigma)\neq (\bj, \sigma')$ }
\end{cases}.
\end{align}
Let $\Lambda_{\mathrm{e}}=\{\bj\in \Lambda\, |\, \mbox{$|\bj|$  is
even}\}$ and let $\Lambda_{\mathrm{o}}=\{\bj\in \Lambda\, |\,
\mbox{$|\bj|$ is odd}\}$.
\begin{define}
{\rm 
A  {\it zigzag transformation } is  a unitary operator defined by
\begin{align}
\mathscr{V}= \prod_{\bj\in \Lambda_{\mathrm{o}}} v_{\bj\uparrow} v_{\bj\downarrow}.
\end{align} 
}
\end{define} 

We remark  that 
\begin{align}
\mathscr{V} c_{\bj\sigma} \mathscr{V}^{-1}
=\begin{cases}
c_{\bj\sigma}^* & \mbox{if $\bj\in \Lambda_{\mathrm{o}}$}\\
c_{\bj\sigma} & \mbox{if $\bj\in \Lambda_{\mathrm{e}}$}
\end{cases} ,
\ \ \ \mathscr{V} q_{\bj} \mathscr{V}^{-1} =(-1)^{|\bj|} q_{\bj}. \label{Zigzag}
\end{align} 

\begin{lemm}\label{ZigzagLemm}
Let $\tilde{H}=\mathscr{V} H \mathscr{V}^{-1}$. We have 
$\tilde{H}=T+W$, where
\begin{align}
T
&= 
\sum_{\la \bi, \bj\ra}
\sum_{\sigma=\uparrow, \downarrow}
(-t) \Big(
c_{\bi\sigma}^* c_{\bj\sigma}^*
+
c_{\bj\sigma}c_{\bi\sigma}
\Big), \label{PP}\\
W&= U\sum_{\bj\in \Lambda} q_{\bj}^2-V\sum_{\la
 \bi, \bj\ra} q_{\bi} q_{\bj}+\frac{\Delta}{2}\sum_{\bj \in \Lambda} q_{\bj}.\label{PP2}
\end{align} 
\end{lemm} 
{\it Proof.} Note  that 
\begin{align}
\sum_{\la \bi, \bj\ra}\sum_{\sigma=\uparrow, \downarrow}\big(
c_{\bi\sigma}^* c_{\bj\sigma}
+c_{\bj\sigma}^* c_{\bi\sigma}
\big)
=\sum_{\bj\in \Lambda_{\mathrm{e}}}\sum_{\sigma= \uparrow, \downarrow}
 \sum_{k=1, 2}
 \sum_{\vepsilon=\pm 1}
\Big(
c_{\bj\sigma}^*
 c_{\bj+\vepsilon {\boldsymbol \delta}_k\sigma}+\mathrm{h.c.}
\Big),
\end{align}
where  ${\boldsymbol \delta}_1=(1, 0)$ and  $
{\boldsymbol \delta}_2=(0, 1)
$.
Thus, by using  Eq. (\ref{Zigzag}), we obtain Eqs. (\ref{PP}) and (\ref{PP2}).  $\Box$

\subsection{General strategy}

By Lemma \ref{ZigzagLemm}, we know that \begin{align}
(-1)^{|\bj|}\la q_{\mb{o}} q_{\bj}\ra_{\beta, \Lambda, H}
=\la q_{\mb{o}} q_{\bj}\ra_{\Lambda}, 
\end{align} 
where $\la \cdot\ra_{\beta, \Lambda,  \tilde{H}} $ is abbreviated  
$\la \cdot\ra_{\Lambda}$.
Thus, the following theorem holds:
\begin{Thm}\label{Trans}
 Equation (\ref{LROClaim}) is equivalent to   
\begin{align}
\liminf_{|\bj|\to \infty} \la q_{\bj}  q_{\mb{o}} \ra>0 \label{EquivExp}
\end{align} 
for sufficiently large $\beta$ and small $t$, where $\la
 \cdot\ra=\lim_{L\to \infty}\la
\cdot \ra_{\Lambda}$.
\end{Thm}

Let $E_{\bj}(\cdot)$ be the spectral measure of
$q_{\bj}$.
We set 
\begin{align}
P_{ \bj}^{(0)}=E_{\bj}(\{0\}),\ \ 
P_{\bj}^{(+)}=E_{\bj}(\{0, +1\}), \ \ \ P_{
 \bj}^{(-)}=E_{\bj}(\{-1\}). \label{DefP}
\end{align} 
\begin{Thm}\label{Strategy}
For all $\bj\in \Lambda, \beta>0$ and $\Lambda\subset \BbbZ^{2}$, 
\begin{align}
\la q_{{\mb o}} q_{\bj}
\ra_{\Lambda} \ge
 1-3 
\Big\la
 P_{\mb{o}}^{(0)}\Big\ra_{\Lambda}
-2\Big\la P_{ \mb{o}}^{(+)}P_{\bj}^{(-)}\Big\ra_{\Lambda}
-2\Big\la P_{ \mb{o}}^{(-)}P_{\bj}^{(+)}\Big\ra_{\Lambda}.\label{FirstStep}
\end{align} 
\end{Thm} 
{\it Proof.}
Let $
P_{\bj}^{\lambda=0}=E_{\bj}(\{0\}), P_{\bj}^{\lambda=+1}=E_{\bj}(\{+1\})
$ and $P_{\bj}^{\lambda=-1}=E_{\bj}(\{-1\})$.
Note that $
P_{\bj}^{(0)}=P_{\bj}^{\lambda=0}, P_{\bj}^{(-)}=P_{\bj}^{\lambda=-1}
$, but $P^{(+)}_{\bj}\neq P^{\lambda=+1}_{\bj}$.
By the spectral theorem, 
\begin{align}
\big\la q_{\mb{o}} q_{\bj}\big\ra_{\Lambda} 
=&\sum_{\lambda, \lambda'\in \{-1, 0,
 1\}} \lambda \lambda' \Big\la E_{{\mb{o}}}(\{\lambda\})
E_{ \bj}(\{\lambda'\})\Big\ra_{\Lambda}\no
= &
\Big\la P_{ \mb{o}}^{\lambda=+1} P_{ \bj}^{\lambda=+1}\Big\ra_{\Lambda} 
+
\Big\la P_{\mb{o}}^{\lambda=-1}
 P_{ \bj}^{\lambda=-1}\Big\ra_{\Lambda}\no
&-
\Big\la P_{ \mb{o}}^{\lambda=+1} P_{ \bj}^{\lambda=-1}\Big\ra_{\Lambda}
 -
\Big\la P_{ \mb{o}}^{\lambda=-1}
 P_{ \bj}^{\lambda=+1}\Big\ra_{\Lambda}.\label{PEq1}
\end{align} 
From
\begin{align}
P_{\bj}^{\lambda=+1} +P_{\bj}^{\lambda=-1}+P_{\bj}^{\lambda=0}=\one, \label{SumOne}
\end{align} 
 it follows that 
\begin{align}
\Big\la P_{\mb{o}}^{\lambda=+1} P_{
 \bj}^{\lambda=+1}\Big\ra_{\Lambda}
&=
\Big\la P_{\mb{o}}^{\lambda=+1}
\Big(\one-P_{\bj}^{\lambda=-1}-P_{\bj}^{\lambda=0}\Big)\Big\ra_{\Lambda}\no
&\ge 
\Big\la P_{\mb{o}}^{\lambda=+1}\Big\ra_{\Lambda}
-\Big\la P_{\mb{o}}^{\lambda=+1}
 P_{\bj}^{\lambda=-1}\Big\ra_{\Lambda}-\Big\la P_{ \mb{o}}^{\lambda=0}\Big\ra_{\Lambda}, \label{PInq1}
\end{align} 
where we have used the fact that 
$
\Big\la P_{\mb{o}}^{\lambda=+1} P_{\bj}^{\lambda=0}\Big\ra_{\Lambda} 
\le \Big\la
P_{\mb{o}}^{\lambda=0}\Big\ra_{\Lambda}
$ (this inequality is an immediate consequence of the Schwartz
inequality).
Similarly, we obtain
\begin{align}
\Big\la P_{\mb{o}}^{\lambda=-1} P_{
 \bj}^{\lambda=-1}\Big\ra_{\Lambda}
\ge 
\Big\la P_{\mb{o}}^{\lambda=-1}\Big\ra_{\Lambda}
-\Big\la P_{\mb{o}}^{\lambda=-1}
 P_{\bj}^{\lambda=+1}\Big\ra_{\Lambda}-\Big\la P_{
 \mb{o}}^{\lambda=0}\Big\ra_{\Lambda}. \label{PInq9}
\end{align} 
From Eq.  (\ref{SumOne}), we see that 
\begin{align}
\Big\la P_{ \mb{o}}^{\lambda=+1} \Big\ra_{\Lambda}
+
\Big\la P_{ \mb{o}}^{\lambda=-1} \Big\ra_{\Lambda}
=
1- \Big\la
 P_{ \bj}^{\lambda=0}\Big\ra_{\Lambda}.\label{PInq10}
\end{align} 
Thus, from Eqs.  (\ref{PInq1})--(\ref{PInq10}),
\begin{align}
&\Big\la P_{ \mb{o}}^{\lambda=+1} P_{ \bj}^{\lambda=+1}\Big\ra_{\Lambda}
+
\Big\la P_{ \mb{o}}^{\lambda=-1} P_{ \bj}^{\lambda=-1}\Big\ra_{\Lambda}\no
\ge& 
1-3 \Big\la P_{
 \mb{o}}^{\lambda=0}\Big\ra_{\Lambda}
-\Big\la P_{\mb{o}}^{\lambda=+1}P_{ \bj}^{\lambda=-1}\Big\ra_{\Lambda}
-\Big\la P_{\mb{o}}^{\lambda=-1}P_{ \bj}^{\lambda=+1}\Big\ra_{\Lambda}. \label{InqInq}
\end{align} 
Inserting Eq. (\ref{InqInq}) into Eq.  (\ref{PEq1}) gives 
\begin{align}
\la q_{{\mb o}} q_{\bj}
\ra_{\Lambda} \ge
 1-3 
\Big\la
 P_{\mb{o}}^{\lambda=0}\Big\ra_{\Lambda}
-2\Big\la P_{ \mb{o}}^{\lambda=+1}P_{\bj}^{\lambda=-1}\Big\ra_{\Lambda}
-2\Big\la P_{ \mb{o}}^{\lambda=-1}P_{\bj}^{\lambda=+1}\Big\ra_{\Lambda}.
\end{align}
Since $
\Big\la P_{{\bf o}}^{\lambda=+1} P_{\bj}^{\lambda=-1}\Big\ra_{\Lambda}
\le 
\Big\la P_{{\bf o}}^{(+)} P_{\bj}^{(-)}\Big\ra_{\Lambda}
$
and 
$
\Big\la P_{{\bf o}}^{\lambda=-1} P_{\bj}^{\lambda=+1}\Big\ra_{\Lambda}
\le 
\Big\la P_{{\bf o}}^{(-)} P_{\bj}^{(+)}\Big\ra_{\Lambda}
$, we conclude (\ref{FirstStep}).
 $\Box$
\medskip\\

Thus, to prove  Eq. (\ref{EquivExp}), it suffices to show the following:
\begin{Thm}\label{Strategy2}
For arbitrary $\vepsilon>0$, there exists $\Lambda_0\subset
 \BbbZ^2, \beta_0>0$ and $t_0\in (0, 1)$ such that if 
$\Lambda \supseteq \Lambda_0, \beta >\beta_0$ and $0<t<t_0$, then 
\begin{itemize}
\item[{\rm {\bf (A)}}] $\Big\la P_{\mb{o}}^{(+)}
	     P_{\bj}^{(-)}\Big\ra_{\Lambda} \le \vepsilon$,  $\Big\la P_{\mb{o}}^{(-)}
	     P_{\bj}^{(+)}\Big\ra_{\Lambda} \le \vepsilon$,
\item[{\rm {\bf (B)}}] $\Big\la P_{\mb{o}}^{(0)}\Big\ra_{\Lambda} \le \vepsilon$.
\end{itemize} 
\end{Thm}

The proof of Theorem \ref{Strategy2} {\bf (A)} is very
complicated,
so  we  present  a concise  strategy to prove it  in Section \ref{STSEC}.
A proof of Theorem \ref{Strategy2} {\bf (B)} appears in Section \ref{Proof(B)}.

\subsection{Strategy of proof of Theorem \ref{Strategy2} (A)}
\label{STSEC}
We only  present a proof of the inequality 
$
\Big\la P_{\mb{o}}^{(+)}
	     P_{\bj}^{(-)}\Big\ra_{\Lambda} \le \vepsilon
$, because the proof of the second  inequality is quite similar.

The proof of Theorem \ref{Strategy2} {\bf (A)} consists of the following steps:
\begin{itemize}
\item[] Step A-1:  Find a key inequality.
\item[] Step A-2: Apply  modified chessboard estimate.
\item[] Step A-3: Estimate $R_{\mathrm{Low}}^{\pm}$ and $R_{\mathrm{High}}^{\pm}$.
\item[] Step A-4: Apply principle   of exponential localization of eigenvectors.
\item[] Step A-5: Complete the proof of  Theorem \ref{Strategy2} {\bf (A)}.
\end{itemize} 
In what follows, we will explain each step.

\subsubsection{Step A-1: Key inequality}
\begin{define}{\rm
 We regard $\Lambda$ as a two-dimensional torus. 
\begin{itemize}
\item The set of all connected sets\footnote{
We say that  a subset $\gamma$ of $\Lambda$ is {\it connected} if any of
      its sites are linked by a path in $\gamma$.
} in $\Lambda$ is denoted 
$\mathscr{S}_{\Lambda}$:
$
\mathscr{S}_{\Lambda}=\{\gamma\subseteq \Lambda
\ |\  \mbox{$\gamma$: connected}
\}
$.
\item By a  {\it contour}, we mean the set $\partial \gamma $ of the nearest neigbor pairs
      associated with the boundary of  a set $\gamma
      \in \mathscr{S}_{\Lambda}$ such that 
\begin{align}
\partial \gamma=\Big\{\la \bi_1, \bj_1\ra,\dots, \la \bi_{\ell}, \bj_{\ell}\ra\, \Big|\,
 \bi_k\in \gamma, \bj_k\notin \gamma \Big\}.
\end{align} 

\end{itemize} 
}
\end{define} 

We present the proof of  the following theorem  in
Section \ref{PeiArg}:
\begin{Thm}\label{Cont}
\begin{align}
\Big\la P^{(+)}_{\bfm} P^{(-)}_{\bn} \Big\ra_{\Lambda} 
\le 
 \sum_{ {\gamma\in \mathscr{S}_{\Lambda}}\atop{
\bfm \in \gamma, \bn\notin\gamma}}
\Bigg\la \prod_{\la \bi, \bj\ra\in \partial \gamma} P_{\bi}^{(+)}P_{\bj}^{(-)}
\Bigg\ra_{\Lambda}. \label{ContEq}
\end{align} 
\end{Thm}

\subsubsection{Step A-2: Modified chessboard estimate}
Set $L=2M+1$.
We define projections $\mb{P}_{\Lambda}^{(+)}$ and
$\mb{P}_{\Lambda}^{(-)}$
by 

\begin{align}
\mathbf{P}_{\Lambda}^{(+)}&=
\Bigg[\prod_{m=1}^M \prod_{n=-L}^{L-1} P_{(-L+4m, n)}^{(+)}P^{(-)}_{(-L+4m+1,
 n)}P^{(-)}_{(-L+4m+2, n)} P_{(-L+4m+3, n)}^{(+)}\Bigg]
\partial \mathbf{P}^{(+)}
, \\
\mathbf{P}_{\Lambda}^{(-)}&=
\Bigg[\prod_{m=1}^M \prod_{n=-L}^{L-1} P_{(-L+4m, n)}^{(-)}P^{(+)}_{(-L+4m+1,
 n)}P^{(+)}_{(-L+4m+2, n)} P_{(-L+4m+3, n)}^{(-)}\Bigg]
\partial \mathbf{P}^{(-)},
\end{align} 
where
\begin{align}
\partial \mathbf{P}^{(\omega)}&=\prod_{n=-L}^{L-1} P_{(L-2,
 n)}^{(\omega)}P_{(L-1, n)}^{(\omega)},\ \ \omega=+, -.
\end{align}

By using  the {\it modified  chessboard estimate}, we present the proof of  the following
theorem 
in Section \ref{PfRP}.

\begin{Thm}\label{MainRP}
Let $\mathcal{P}_{\Lambda}=\max\big\{\big\la \mb{P}_{\Lambda}^{(+)}\big\ra_{\Lambda}, 
\big\la \mb{P}_{\Lambda}^{(-)}
\big\ra_{\Lambda} \big\}$. 
This gives
\begin{align}
\Bigg\la \prod_{\la \bi, \bj\ra\in \partial \gamma} P_{\bi}^{(+)}P_{\bj}^{(-)}
\Bigg\ra_{\Lambda}\le 
\mathcal{P}_{\Lambda}^{|\partial\gamma|/2|\Lambda|}.
\end{align} 
\end{Thm}

\begin{coro}\label{Peierls}
There exists a  $C>0$  such that 
\begin{align}
\Big\la P_{\bfm}^{(+)}P_{\bn}^{(-)}\Big\ra_{\Lambda}\le C\sum_{\ell=4}^{\infty} \ell^2
 3^{\ell} \mathcal{ P}_{\Lambda}^{\ell/2|\Lambda|}. \label{ChessCoro}
\end{align} 
\end{coro} 
{\it Proof.}
By  Theorems \ref{Cont} and \ref{MainRP}, 
\begin{align}
\Big\la P^{(+)}_{{\boldsymbol m}} P^{(-)}_{{\boldsymbol n}} 
\Big\ra_{\Lambda} 
\le 
 \sum_{ {\gamma\in \mathscr{S}_{\Lambda}}\atop{
{\boldsymbol m}\in \gamma, {\boldsymbol n}\notin\gamma}}
\mathcal{P}_{\Lambda}^{|\partial \gamma|/2|\Lambda|}, \label{ContEq2}
\end{align} 
where $|\partial \gamma|$ is  the number of  nearest neighbor
pairs in $\partial \gamma$.
Because the smallest contours have surface $4$, we have
\begin{align}
\mbox{RHS of (\ref{ContEq2})}
=\sum_{\ell=4}^{\infty} \#\Big\{
\gamma\in \mathscr{S}_{\Lambda}\, \Big|\, {\boldsymbol m}\in \gamma,\,
 {\boldsymbol n}\notin\gamma,\
 |\partial \gamma|=\ell
\Big\}
\mathcal{P}_{\Lambda}^{\ell/2|\Lambda|}.
\end{align} 
By the standard Peierls  argument, there exists a  $C>0$
 such that\footnote{
The factor $3^{\ell}$ comes from the fact that the number of
connected surfaces $N$ consisting of  $\ell$ blocks and containing a fixed block
 is bounded above by $3^{\ell}:\, N\le 3^{\ell}$.
The factor $C\ell^2$ comes from the fact that each $\gamma$ must contain
${\boldsymbol m}$. 
}  
\begin{align}
\#\Big\{
\gamma\in \mathscr{S}_{\Lambda}\, \Big|\, {\boldsymbol m}\in \gamma,\, 
{\boldsymbol n}\notin\gamma,\
 |\partial \gamma|=\ell
\Big\}\le C \ell^2 3^{\ell}.
\end{align} 
Thus, the assertion in Corollary \ref{Peierls} holds. $\Box$

\subsubsection{Step A-3: Estimate $R_{\mathrm{Low}}^{\pm}$ and $R_{\mathrm{High}}^{\pm}$}
To prove Theorem \ref{Strategy2} {\bf(A)},  showing that 
the right-hand side of Eq.  (\ref{ChessCoro}) $\le  \vepsilon$
suffices.

Let $E(\cdot)$ be the spectral measure of $\tilde{H}$.
For each $\delta>0$, we use
\begin{align}
E_{\delta}=E_{\tilde{H}}\big([\underline{e},\
 \underline{e}+\delta|\Lambda|]\big),\ \ \
 E_{\delta}^{\perp}=\one-E_{\delta}, \label{DefE}
\end{align} 
where $\underline{e}=\min \mathrm{spec}(\tilde{H})$ (i.e.,  the ground
state energy).
The choice of the quantity
$\delta$ will be addressed  later.

We divide $\Big\la \mb{P}_{\Lambda}^{(\pm)}\Big\ra_{\Lambda}$ into two pieces: $
\Big\la \mb{P}_{\Lambda}^{(\pm)}\Big\ra_{\Lambda}=R_{\mathrm{Low}}^{(\pm)}+R_{\mathrm{High}}^{(\pm)}
$, where
\begin{align}
R_{\mr{Low}}^{(\pm)} =\Big\la E_{\delta} \mb{P}_{\Lambda}^{(\pm)}\Big\ra_{\Lambda},\ \ \ 
R_{\mr{High}}^{(\pm)}= \Big\la E_{\delta}^{\perp} \mb{P}_{\Lambda}^{(\pm)}\Big\ra_{\Lambda}.
\end{align}

\begin{Thm}\label{EstR1}
We assert  the following:
\begin{itemize}
\item[{\rm (i)}]
$
\big|
R_{\mathrm{Low}}^{(\omega)}
\big|
\le  2^{|\Lambda|}
\big\{\Tr_{\mathfrak{H}}
\big[
\mb{P}_{\Lambda}^{(\omega)} E_{\delta} \big]
\big\}^{1/2}
$ for each  $\omega=+, -$.
\item[{\rm (ii)}]
$
\big|
R_{\mathrm{High}}^{(\omega)}
\big|
\le
4^{|\Lambda|}
e^{-\beta \delta |\Lambda|}
$ for each $\omega=+, -$.
\end{itemize} 
\end{Thm} 
{\it Proof.}
(i)
By  the Schwartz inequality, 
we have
\begin{align}
\Big|
\Tr_{\mathfrak{H}}
\Big[E_{\delta}\mb{P}_{\Lambda}^{(\omega)} e^{-\beta \tilde{H}}
\Big]
\Big|
\le 
\big\| e^{-\beta \tilde{H}}\big\|_{\mathrm{HS}} \big\| E_{\delta} \mb{P}_{\Lambda}^{(\omega)}\big\|_{\mathrm{HS}}
\le e^{-\beta \underline{e}} 2^{|\Lambda|} 
\Big\{\Tr_{\mathfrak{H}}\Big[\mb{P}_{\Lambda}^{(\omega)} E_{\delta}\Big]\Big\}^{1/2},
\end{align}
where $\|A\|_{\mathrm{HS}}:=\big\{
\Tr_{\mathfrak{H}}[|A|^2]
\big\}^{1/2}$, the Hilbert--Schmidt norm.
Because  $Z_{\Lambda}(\beta)=\Tr_{\mathfrak{H}}[e^{-\beta \tilde{H}}]
\ge e^{-\beta \underline{e}}$, we conclude part (i) of Theorem \ref{EstR1}.

(ii)
We observe that 
\begin{align}
\Big|
R_{\mathrm{High}}^{(\omega)}
\Big|
=\Bigg|
\Tr_{\mathfrak{H}} \Big[
E_{\delta}^{\perp} \mathbf{P}_{\Lambda}^{(\omega)}
e^{-\beta \tilde{H}}
\Big]\Big/Z_{\Lambda}(\beta)
\Bigg|
\le e^{-\beta(\underline{e}+\delta |\Lambda|)} 4^{|\Lambda|}
 e^{+\beta\underline{e}}=4^{|\Lambda|} e^{-\beta \delta |\Lambda|}.
\end{align} 
This completes the proof. $\Box$

\subsubsection{Step A-4: Exponential localization of eigenvectors}

By using  the {\it principle of exponential localization} \cite{FL}, we  show
the following in Section \ref{PfLocal}:

\begin{Thm}\label{LocEstRep}
Set $S=2V-U$. Let 
\begin{align}
\mathscr{J}=
\begin{cases}
\frac{1}{4}(\Delta +V) & \mbox{if $S\ge 0$}\\
\frac{1}{2}(S+\frac{\Delta}{2})+\frac{V}{4} & \mbox{if $S<0$}.
\end{cases} 
\label{DefJ}
\end{align} 
We choose $\delta $ as $\delta=\beta^{-\xi}$ with
 $\xi\in (0, 1)$.  If $\displaystyle \frac{\Delta}{2}+S>0$, then
\begin{align}
\Tr_{\mathfrak{H}}\Big[\mb{P}^{(\pm)}_{ \Lambda} E_{\delta}\Big] \le
4^{|\Lambda|} \gamma^{d},
\end{align} 
where $\gamma$ and $d$ satisfy
\begin{align}
\gamma=& t\Big(
1+8
\mathscr{J}^{-1}
\eta^{-1}
\Big)+\mathcal{O}\big(\beta^{-\xi}\big)
,\ \ \ \\
d>&
\frac{1-\eta
}{G} \mathscr{J} |\Lambda|+
\mathcal{O}\big(|\Lambda|^{1/2}\big)
\end{align}
with $G=6(|S|+V)+\Delta$ and $\eta\in (0, 1)$.
\end{Thm}
\begin{rem}
{\rm 
Because  $\frac{\Delta}{2}+S>0$,  it holds that $\mathscr{J}>0$. $\diamondsuit$
}
\end{rem}

\subsubsection{Step A-5: Completion of  the proof of Theorem \ref{Strategy2} {\bf (A)}.}
By Theorems \ref{EstR1} and \ref{LocEstRep}, we have
\begin{align}
\Big \la \mathbf{P}_{\Lambda}^{(\pm)} \Big\ra_{\Lambda} \le 4^{|\Lambda|}
 \big(e^{-\mathscr{A} |\Lambda|}
+e^{-\beta \delta |\Lambda|}
\big),
\end{align} 
where 
\begin{align}
\mathscr{A}=\Bigg\{\frac{1-\eta}{2G}
\mathscr{J}
+\mathcal{O}(|\Lambda|^{-1/2})
\Bigg\}\log t^{-1}\Bigg\{
1+8\mathscr{J}^{-1} \eta^{-1}+\mathcal{O}(\beta^{-\xi})
\Bigg\}.
\end{align} 
Note  that $\mathscr{A}>0$, provided that $t$ is sufficiently small.
 Let 
 $\mathscr{D}=\min\{\mathscr{A}, \beta \delta\}$.
Recall that  $\delta=\beta^{-\xi}$.
We obtain 
\begin{align}
\mathcal{P}_{\Lambda} \le 2\cdot 4^{|\Lambda|} e^{-\mathscr{D}|\Lambda|}. \label{PEST}
\end{align} 
Thus, by Corollary \ref{Peierls}, 
\begin{align}
\Big\la P_{\bf o}^{(+)} P_{\bj}^{(-)}\Big\ra_{\Lambda}
\le C \sum_{\ell=4}^{\infty} \ell^2 (24)^{\ell} \exp\bigg\{-\frac{\mathscr{D}}{2}\ell \bigg\}.
\end{align} 
Because we can choose $\mathscr{D}$ as large as we wish by using a 
sufficiently large $\beta $ and    sufficiently small $t$, 
we obtain Theorem \ref{Strategy2} {\bf (A)}. $\Box$

\section{Proof of Theorem \ref{Cont}}\label{PeiArg}
Although  Theorem \ref{Cont} is proven  in  Ref. \cite{FL}, 
we provide its  proof  here  for the  reader\rq{}s convenience.

\setcounter{equation}{0}
\begin{define}{\rm
\begin{itemize}
\item A {\it configuration} $c$ is a function on $\Lambda$ with values in
      $\{+, -\}$ such that $c({\boldsymbol m})=+$ and $c({\boldsymbol n})=-$.
We denote by $\mathcal{C}$ the set of all configurations on $\Lambda$.
\item For each $c \in \mathcal{C}$, we set
\begin{align}
\Gamma(c)
=\Big\{\partial \gamma
=&\{\la \bi_1, \bj_1\ra,\dots, \la \bi_{\ell}, \bj_{\ell}\ra\}: \, \mathrm{contour}\,  
\Big|
c(\bi_k)=+,\,   
c(\bj_k)=-, \,   k=1, \dots, \ell
\Big\}.
\end{align} 
Note that the definition of $\Gamma(c)$ is meaningful because of  Remark
      \ref{MinRem} below. $\diamondsuit$
\end{itemize} 
}
\end{define}

\begin{rem}\label{MinRem}
{\rm
For all $c\in \mathcal{C}$, there exists a unique smallest set  $\gamma(c)\in \mathscr{S}_{\Lambda}$ such that 
\begin{itemize}
\item ${\boldsymbol m}\in \gamma(c)$,
\item $c(\bi)=+$ for all $\bi\in \gamma(c)$.
\end{itemize} 
}
\end{rem}

\begin{lemm}\label{Use1}
\begin{align}
\mathcal{C}=\bigcup_{\gamma\in \mathscr{S}_{\Lambda}}
\Big\{
c\in \mathcal{C}
\Big|\, \partial \gamma=\partial \gamma(c)
\Big\}\label{Cex}
,
\end{align} 
 where $\gamma(c)$ is given in Remark \ref{MinRem}. 
\end{lemm} 
{\it Proof.} We denote by $\mathcal{K}$ the right-hand side of
Eq. 
(\ref{Cex}).  Because $\mathcal{C} \supseteq
\mathcal{K}$ is easy to see,   we  show the converse.
For arbitrarily fixed $c_0\in \mathcal{C}$, we set $\gamma_0=\gamma(c_0)$,
where $\gamma(c)$ is given in Remark \ref{MinRem}.
Showing that $c_0\in \{c\in \mathcal{C}\, |\,
\partial\gamma_0= \partial \gamma(c)\}$ is then trivial.
Thus, $c_0\in \mathcal{K}$, which implies $\mathcal{C}\subseteq
\mathcal{K}$. $\Box$

\begin{flushleft}
{\it Completion of proof of Theorem  \ref{Cont}}
\end{flushleft} 
Since $P_{\bj}^{(+)} + P_{\bj}^{(-)}=\one$, we have
\begin{align}
\Big\la P_{\boldsymbol m}^{(+)}P_{\boldsymbol n}^{(-)}\Big\ra_{\Lambda}
=&\Bigg\la
P_{{\boldsymbol m}}^{(+)}P_{\boldsymbol n}^{(-)} \prod_{{\bj\in
 \Lambda}\atop{\bj\neq {\boldsymbol m},{\boldsymbol n}}}\Big[P_{\bj}^{(+)}+P_{\bj}^{(-)}\Big]
\Bigg\ra_{\Lambda}\no
=& \sum_{c\in \mathcal{C}}\Bigg\la \prod_{\bj\in \Lambda}
 P_{\bj}^{(c(\bj))}\Bigg\ra_{\Lambda}\no
=&
 \sum_{\gamma\in \mathscr{S}_{\Lambda}} \sum_{c\in \{c|\partial
 \gamma=\partial \gamma(c)\}} \Bigg\la \prod_{\bj\in \Lambda}
 P_{\bj}^{(c(\bj))}\Bigg\ra_{\Lambda}\ \ (\because \mbox{Lemma \ref{Use1}}) \no
<& \sum_{\gamma\in \mathscr{S}_{\Lambda}} \sum_{c\in \{c|\partial
 \gamma\in \Gamma(c)\}} \Bigg\la \prod_{\bj\in \Lambda}
 P_{\bj}^{(c(\bj))}\Bigg\ra_{\Lambda}\ \ (\because 
\{c|\partial
 \gamma=\partial \gamma(c)\} \subset \{c|\partial
 \gamma\in \Gamma(c)\}
) \no
=& \sum_{\gamma\in \mathscr{S}_{\Lambda}} \sum_{c\in \{c|\partial
 \gamma\in\Gamma(c)\}}
\Bigg\la
\Bigg[
\prod_{\la \bi, \bj\ra\in \partial \gamma} P_{\bi}^{(+)}P_{\bj}^{(-)}
\Bigg]
\Bigg[
\prod_{\bj\in \Lambda_{\partial \gamma}} P_{\bj}^{(c(\bj))}
\Bigg]
P_{\boldsymbol m}^{(+)} P_{\boldsymbol n}^{(-)}
\Bigg\ra_{\Lambda}, \label{PeiInq}
\end{align} 
where $\Lambda_{\partial \gamma}$ is defined as follows:
For each $\partial \gamma=\{\la \bi_1, \bj_1\ra, \dots, \la \bi_{\ell},
\bj_{\ell}\ra\}$, let
$[\partial \gamma]=\{\bi_k\}_{k=1}^{\ell}\cup \{\bj_k\}_{k=1}^{\ell}
\subseteq \Lambda$. Then $\Lambda_{\partial \gamma}$ is given by 
$
\Lambda_{\partial \gamma}=\Lambda\backslash \{[\partial \gamma] \cup
\{{\boldsymbol m},{\boldsymbol  n }\} \}
$.
Note  the following fact:
\begin{align}
\sum_{c\in \{c|\partial \gamma \in \Gamma(c)\}} \prod_{\bj\in
 \Lambda_{\partial \gamma}} P_{\bj}^{(c(\bj))}
=\prod_{\bj\in \Lambda_{\partial\gamma}}\Big[P_{\bj}^{(+)}+P_{\bj}^{(-)}\Big]
=
\one.
\end{align} 
Thus,  because $P_{\boldsymbol m}^{(+)}P_{\boldsymbol n}^{(-)}\le \one$, 
we have
\begin{align}
\mbox{RHS of (\ref{PeiInq})}
=\sum_{\gamma \in \mathscr{S}_{\Lambda} }
\Bigg\la
\Bigg[
\prod_{\la \bi, \bj\ra\in \partial \gamma} P_{\bi}^{(+)}P_{\bj}^{(-)}
\Bigg]
P_{\boldsymbol m}^{(+)} P_{\boldsymbol n}^{(-)}
\Bigg\ra_{\Lambda}
\le \sum_{\gamma \in \mathscr{S}_{\Lambda} }
\Bigg\la
\prod_{\la \bi, \bj\ra\in \partial \gamma} P_{\bi}^{(+)}P_{\bj}^{(-)}
\Bigg\ra_{\Lambda}.
\end{align}
This completes the proof. $\Box$

\section{Proof of Theorem \ref{MainRP}} \label{PfRP}
\setcounter{equation}{0}

\subsection{Reflection positivity}
We use reflection positivity.

Let $\Lambda_L=\{\bj=(j_1, j_2)\in \Lambda\, |\, j_1\le -1\}$ and $\Lambda_R
=\{\bj=(j_1, j_2)\in \Lambda\, |\, j_1\ge 0\}$. Since $\ell^2(\Lambda)=
\ell^2(\Lambda_L) \oplus \ell^2(\Lambda_R)$, 
we have
\begin{align}
\ell^2(\Lambda) \oplus \ell^2(\Lambda)
=\big(
\ell^2(\Lambda_L) \oplus \ell^2(\Lambda_L)
\big)
\oplus 
\big(
\ell^2(\Lambda_R) \oplus \ell^2(\Lambda_R)
\big).
\end{align} 
Thus, we obtain the following identification:
\begin{align}
\mathfrak{H}=\mathfrak{H}_L\otimes \mathfrak{H}_R, \label{HilLR}
\end{align} 
where $\mathfrak{H}_L=\Fock(\ell^2(\Lambda_L)\oplus \ell^2(\Lambda_L))$
 and $\mathfrak{H}_R=
\Fock(\ell^2(\Lambda_R)\oplus \ell^2(\Lambda_R))
$. 
Note that, by applying Eq.  (\ref{HilLR}), the annihilation operator
 can be  expressed as  
\begin{align}
c_{\bj\sigma}=
\begin{cases}
c_{\bj\sigma}\otimes \one & \bj \in \Lambda_L\\
(-1)^{N_L}\otimes c_{\bj \sigma} & \bj \in \Lambda_R
\end{cases},\label{TensorC}
\end{align} 
where $N_L=\sum_{\bj \in \Lambda_L} n_{\bj}$.

\begin{lemm}
According to Eq.  (\ref{HilLR}), 
we have the following:
\begin{itemize}
\item[{\rm (i)}]
$
T=T_L\otimes \one+\one \otimes T_R+T_{LR}
$, where 
\begin{align}
T_L=&
\sum_{\bj\in \Lambda_{\mathrm{e}},\  j_1\le -2}\sum_{\sigma=\uparrow, \downarrow}
\sum_{k=1, 2} \sum_{\vepsilon=\pm}'
(-t)  \Big(
c_{\bj \sigma}^* c_{\bj+\vepsilon {\boldsymbol \delta}_k\sigma}^*
+
\mathrm{h. c. }
\Big), \label{T1}\\
T_R =&\sum_{\bj\in \Lambda_{\mathrm{e}},\  j_1\ge 0}
\sum_{\sigma=\uparrow, \downarrow}
\sum_{k=1, 2} \sum_{\vepsilon=\pm}''
(-t)  \Big(
c_{\bj\sigma}^* c_{\bj+\vepsilon {\boldsymbol \delta}_k\sigma}^*
+
\mathrm{h. c. }
\Big),\label{TRc}\\
T_{LR}=&
\sum_{\bj\in \Lambda_{\mathrm{e}},\ j_1=0} \sum_{\sigma=\uparrow, \downarrow}(-t)
\Bigg\{
\Big[
(-1)^{N_L} c_{\bj-{\boldsymbol \delta}_1  \sigma}
^*\Big]
\otimes
c_{\bj\sigma}^*
+\mathrm{h.c.}
\Bigg\}\no
&+ 
\sum_{\bj \in \Lambda_{\mathrm{e}},\ j_1=L-1} \sum_{\sigma=\uparrow, \downarrow}(-t)
\Bigg\{
\Big[
(-1)^{N_L} c_{\bj +{\boldsymbol \delta}_1  \sigma}
^*\Big]
\otimes
c_{\bj \sigma}^*
+\mathrm{h.c.}
\Bigg\}.\label{T2}
\end{align} 
Here, 
 $\displaystyle \sum_{\vepsilon=\pm}'$ refers to a sum over pairs
 $\la \bj,  \bj+\vepsilon {\boldsymbol \delta}_k\ra$ such that $\bj,
	     \bj+\vepsilon {\boldsymbol \delta}_k\in
 \Lambda_L$. Similarly,   $\displaystyle \sum_{\vepsilon=\pm}''$ refers to a sum over pairs
 $\la \bj, \bj+\vepsilon {\boldsymbol \delta}_k\ra$ such that $\bj,
	     \bj+\vepsilon {\boldsymbol \delta}_k\in
 \Lambda_R$.
\item[{\rm (ii)}]
 $W=W_L\otimes \one +\one \otimes W_R+W_{LR}$, where  
\begin{align}
W_L=& -S\sum_{\bj\in \Lambda_L}
 q_{\bj }^2+\frac{V}{2}\sum_{\la \bi,  \bj\ra,\ \bi, \bj\in \Lambda_L} (q_{\bi}- q_{\bj})^2,\\
 W_R =& 
-S\sum_{\bj\in \Lambda_R} q_{\bj}^2+\frac{V}{2}\sum_{\la
 \bi; \bj\ra,\ \bi, \bj\in \Lambda_R} (q_{\bi}- q_{\bj})^2,\\
W_{LR}=&
-V \sum_{\bj\in \Lambda_{\mathrm{e}},\ j_1=0} q_{\bj-{\boldsymbol \delta}_1}
\otimes q_{\bj}
-V \sum_{\bj \in \Lambda_{\mathrm{e}},\ j_1=L-1} q_{\bj+{\boldsymbol \delta}_1}\otimes q_{\bj}.
\end{align} 
\end{itemize} 
\end{lemm} 
{\it Proof.} (i)
Note 
\begin{align}
T=
\sum_{\bj\in \Lambda_{\mathrm{e}}}\sum_{\sigma=\uparrow, \downarrow}
\sum_{k=1, 2} \sum_{\vepsilon=\pm}
(-t)  \Big(
c_{\bj \sigma}^* c_{\bj+\vepsilon {\boldsymbol \delta}_k\sigma}^*
+
\mathrm{h. c. }
\Big).
\end{align} 
Thus, Eqs.  (\ref{T1})--(\ref{T2}) are easily verified by Eq. 
(\ref{TensorC}). To show (ii), we note  that $W$ can be expressed as 
\begin{align}
W=-S\sum_{\bj\in \Lambda} q_{\bj}^2
+\frac{V}{2}\sum_{\la \bi, \bj\ra}
 (q_{\bi}-q_{\bj})^2+\frac{\Delta}{2}\sum_{\bj\in \Lambda} q_{\bj} \label{Wex}
.\ \ \ \Box
\end{align} 
\medskip\\

For all $\bj\in \Lambda_L$,  we define
\begin{align}
a_{\bj\sigma}=c_{\bj\sigma} (-1)^{N_L}.
\end{align} 

In terms of $a_{\bj\sigma}$, $T_L$ and $T_{LR}$ can be expressed as
follows
(note that  we will  use the fact $c_{\bj+\vepsilon{\boldsymbol \delta}_k}^*
c_{\bj\sigma}^*
=-a_{\bj+\vepsilon{\boldsymbol \delta}_k}^*
a_{\bj\sigma}^*
$).  
\begin{Prop}
We obtain 
\begin{align}
T_L =&\sum_{\bj\in \Lambda_{\mathrm{e}},\  j_1\le -2}
\sum_{\sigma=\uparrow, \downarrow}\sum_{k=1}^{\nu} \sum_{\vepsilon=\pm}'
(+t)  \Big(
a_{\bj\sigma}^* a_{\bj+\vepsilon {\boldsymbol \delta}_k\sigma}^*
+
\mathrm{h. c. }
\Big),\\
T_{LR}=&
\sum_{\bj\in \Lambda_{\mathrm{e}},\ j_1=0} \sum_{\sigma=\uparrow, \downarrow}(-t)
\Big(
a_{\bj-{\boldsymbol \delta}_1 \sigma}^*
\otimes
c_{\bj\sigma}^*
+\mathrm{h.c.}
\Big)\no
&+ 
\sum_{\bj \in \Lambda_{\mathrm{e}},\ j_1=L-1} \sum_{\sigma=\uparrow, \downarrow}(-t)
\Big(
a_{\bj+{\boldsymbol \delta}_1\sigma}^*
\otimes c_{\bj \sigma}^*
+\mathrm{h.c.}
\Big).
\end{align} 

\end{Prop} 
\begin{rem}
{\rm 
Because  $q_{\bj}=\sum_{\sigma=\uparrow, \downarrow}
 a_{\bj\sigma}^*a_{\bj\sigma}-\one$,  the expressions for 
 $W_L$ and
 $W_{LR}$ are  unchanged if we rewrite them in terms of
 $a_{\bj\sigma}$. $\diamondsuit$
}
\end{rem} 

The {\it reflection map} $r_v: \Lambda_R\to \Lambda_L$ is defined by
\begin{align}
r_v(\bj)=(-j_1-1, j_2),\ \ \bj=(j_1, j_2)\in \Lambda_R. \label{thetav}
\end{align} 
Let $\vartheta_v$ be an antiunitary transformation\footnote{
Namely, $\vartheta_v$ is a bijective antilinear map that  satisfies
$
\la \vartheta_v \vphi|\vartheta_v \psi\ra=\overline{
\la \vphi|\psi\ra
}$ for all $\vphi, \psi\in \mathfrak{H}_L$
.
} from $\mathfrak{H}_L$
to $\mathfrak{H}_R$ such that 
\begin{align}
\vartheta_v \Omega_L=\Omega_R,\ \ c_{\bj\sigma}=\vartheta_v a_{r_v(\bj)\sigma}
 \vartheta_v^{-1},\ \ \bj\in \Lambda_R, \label{RPmap}
\end{align} 
where $\Omega_L$ ($\Omega_R$) is the Fock vacuum  in
$\mathfrak{H}_L$ ($\mathfrak{H}_R$).

\begin{lemm}\label{TensorRep}
\begin{itemize}
\item[{\rm (i)}] $T_R=\vartheta_v T_L \vartheta_v^{-1}$.
\item[{\rm (ii)}] \begin{align}
T_{LR}=&\sum_{\bj \in \Lambda_{\mathrm{e}},\ j_1=0}
 \sum_{\sigma=\uparrow, \downarrow}(-t)
\Big(
a_{\bj-{\boldsymbol \delta}_1 \sigma}^*
\otimes 
\vartheta_v 
a_{\bj-{\boldsymbol \delta}_1\sigma}^*
\vartheta_v^{-1}+\mathrm{h.c.}
\Big)\no
&+ 
\sum_{\bj \in \Lambda_{\mathrm{e}},\ j_1=L-1} \sum_{\sigma=\uparrow, \downarrow}(-t)
\Big(
a_{\bj+{\boldsymbol \delta}_1\sigma}^*
\otimes
 \vartheta_v a_{\bj+{\boldsymbol \delta}_1\sigma}^*
\vartheta_v^{-1}
+\mathrm{h.c.}
\Big).
\end{align} 
\item[{\rm (iii)}] $W_R=\vartheta_v W_L\vartheta_v^{-1}$.
\item[{\rm (iv)}]
\begin{align}
W_{LR}=&
-V \sum_{\bj \in \Lambda_{\mathrm{e}},\ j_1=0} q_{\bj-{\boldsymbol \delta}_1}
\otimes \vartheta_v q_{\bj-{\boldsymbol \delta}_1}\vartheta_v^{-1}
-V \sum_{\bj \in \Lambda_{\mathrm{e}},\ j_1=L-1} q_{\bj+{\boldsymbol
 \delta}_1}\otimes \vartheta_v q_{\bj+{\boldsymbol \delta}_1}\vartheta_v^{-1}.
\end{align} 
\end{itemize} 
\end{lemm} 
{\it Proof.} Items (ii), (iii),  and (iv) are easy to check.
To verify  item (i),  note    that $T_L$ can be expressed as 
\begin{align}
T_L=\sum_{\bj \in \Lambda_{\mathrm{o}},\  j_1\le -1}
\sum_{\sigma=\uparrow, \downarrow}
\sum_{k=1,2} \sum_{\vepsilon=\pm}'
(+t)  \Big(
 a_{\bj +\vepsilon {\boldsymbol \delta}_k\sigma}^* a_{\bj\sigma}^*
+
\mathrm{h. c. }
\Big).
\end{align}
Thus, by Eqs. (\ref{TRc}) and (\ref{RPmap}), we obtain  
\begin{align}
T_R &=
\vartheta_v \sum_{\bj\in \Lambda_{\mathrm{e}},\  j_1\ge 0}
\sum_{\sigma=\uparrow, \downarrow}
\sum_{k=1,2} \sum_{\vepsilon=\pm}'
(-t)  \Big(
a_{r_v(\bj)\sigma}^* a_{r_v(\bj+\vepsilon {\boldsymbol \delta}_k)\sigma}^*
+
\mathrm{h. c. }
\Big) \vartheta_v^{-1}\no
&=\vartheta_v 
\sum_{X=(X_1, X_2)\in \Lambda_{\mathrm{o}},\  X_1\le -1}
\sum_{\sigma=\uparrow, \downarrow}\sum_{k=1,2} \sum_{\vepsilon=\pm}'
(-t)  \Big(
a_{X\sigma}^* a_{X +\vepsilon {\boldsymbol \delta}_k\sigma}^*
+
\mathrm{h. c. }
\Big) \vartheta_v^{-1}\no
&=\vartheta_v 
\sum_{X\in \Lambda_{\mathrm{o}},\  X_1\le -1}
\sum_{\sigma=\uparrow, \downarrow}\sum_{k=1,2} \sum_{\vepsilon=\pm}'
(+t)  \Big(
 a_{X +\vepsilon {\boldsymbol \delta}_k\sigma}^* a_{X\sigma}^*
+
\mathrm{h. c. }
\Big) \vartheta_v^{-1}\no
&=\vartheta_v T_L\vartheta_v^{-1}.
\end{align} 
Here, we  have used the fact that  if $\bj\in \Lambda_{\mathrm{e}}$, then $r_v(\bj) \in
\Lambda_{\mathrm{o}}$. $\Box$
\medskip\\

By Theorem \ref{DLSCl} and Lemma \ref{TensorRep}, we immediately obtain the following:
\begin{Thm}\label{RPInq}
Let $A, B\in \mathfrak{B}(\mathfrak{H}_L)$, where
 $\mathfrak{B}(\mathfrak{X})$ is the set of all linear operators on $\mathfrak{X}$
.  The following then holds:
\begin{itemize}
\item[{\rm (i)}] $\displaystyle
\la A\otimes \vartheta_v A\vartheta_v^{-1}\ra_{\Lambda}\ge 0,
$
\item[{\rm (ii)}]
$\displaystyle 
\Big|
\big\la A\otimes \vartheta_v B\vartheta_v^{-1} \big\ra_{\Lambda}
\Big|^2
\le
\big\la A\otimes \vartheta_v A\vartheta_v^{-1}\big\ra_{\Lambda}
 \big\la B\otimes \vartheta_v B\vartheta_v^{-1}\big\ra_{\Lambda}.
$
\end{itemize} 
\end{Thm} 

Theorem \ref{RPInq} is reflection positivity associated with a {\it vertical}
line $j_1=-\frac{1}{2}$.
We can also construct reflection positivity associated with a {\it  horizontal}
line $j_2=-\frac{1}{2}$.

Let $\Lambda^U=\{\bj=(j_1, j_2)\in \Lambda\, |\, j_2\ge 0\}$
and $\Lambda^L=
\{\bj=(j_1, j_2)\in \Lambda\, |\, j_2\le -1\}
$.
As before, we obtain the following identification:
\begin{align}
\mathfrak{H}=\mathfrak{H}^L\otimes \mathfrak{H}^U,
\end{align} 
where $\mathfrak{H}^L=\Fock(\ell^2(\Lambda^L) \oplus \ell^2(\Lambda^L))$
and $\mathfrak{H}^U=\Fock(\ell^2(\Lambda^U)\oplus \ell^2(\Lambda^U))$.
The reflection map $r_h: \Lambda^L\to \Lambda^U$ is given by 
\begin{align}
r_h(\bj)=(j_1, -j_2-1),\ \ \ \bj=(j_1, j_2) \in \Lambda^L.
\end{align} 
Let $\vartheta_h$ be an antiunitary transformation from $\mathfrak{H}^L$
to $\mathfrak{H}^U $ such that
\begin{align}
\vartheta_h\Omega^L=\Omega^U,\ \ c_{\bj\sigma}=\vartheta_h
 a_{r_h(\bj)\sigma} \vartheta_h^{-1}, \ \ \bj\in \Lambda^U, \label{thetah}
\end{align} 
where $\Omega^L$ ($\Omega^U$) is the Fock vacuum in
$\mathfrak{H}^L$ ($\mathfrak{H}^U$).
By a parallel argument, we can prove the following theorem:

\begin{Thm}\label{RPInq2}
Let $A, B\in \mathfrak{B}(\mathfrak{H}^L)$. Then the following holds:
\begin{itemize}
\item[{\rm (i)}] $\displaystyle 
\la A\otimes \vartheta_h A\vartheta_h^{-1}\big\ra_{\Lambda}\ge 0,
$
\item[{\rm (ii)}]
$\displaystyle 
\Big|
\big\la A\otimes \vartheta_h B\vartheta_h^{-1} \big\ra_{\Lambda}
\Big|^2
\le
\big\la A\otimes \vartheta_h A\vartheta_h^{-1}\big\ra_{\Lambda}
 \big\la B\otimes \vartheta_h B\vartheta_h^{-1}\big\ra_{\Lambda}.
$
\end{itemize} 
\end{Thm}

\subsection{Modified chessboard estimate}

To prove Theorem \ref{MainRP}, the chessboard estimate
established from  \cite{FL} can be employed. Unfortunately, the idea of
Ref.  \cite{FL} cannot be
directly applied  because it 
 works well only if $L$ is {\it even}. However, 
our argument of reflection positivity requires that $L$ be
{\it odd}.
Therefore, it is necessary  to extend the chessboard estimate to
overcome this difficulty.

First, we recall the original chessboard estimate.

\begin{Thm}\label{Standard}
Let $\mathfrak{A}$ be a vector space with antilinear involution $J$.
Let $\omega$ be a multilinear functional on $\mathfrak{A}^{2L}$. Assume the following:
\begin{itemize}
\item[{\rm (i)}] $\omega(A_1, \dots, A_{2L})=\omega(A_2, \dots,
	     A_{2L}, A_1)$.
\item[{\rm (ii)}] $
\omega(A_1, \dots, A_L, JA_L, \dots, JA_1)\ge 0.
$ 
\item[{\rm (iii)}] 
\begin{align}
\Big|
\omega(A_1, \dots, A_{2L})
\Big|
\le& \omega\big(A_1, \dots, A_L,  JA_L, \dots, JA_1\big)^{1/2}\no
&\times \omega\big(JA_{2L}, \dots, JA_{L+1}, A_{L+1}, \dots, A_{2L}\big)^{1/2}.
\end{align} 
\end{itemize} 
Then   the following holds:
\begin{align}
\Big|
\omega(A_1, \dots, A_{2L})
\Big|&\le \prod_{j=1}^{2L}
\omega\big(
JA_j, A_j, \dots, JA_j, A_j
\big)^{1/2L}. \label{Chess1}
\end{align} 
\end{Thm}

The following is a modified version of Theorem \ref{Standard}.
In the present work, we  use both Theorems \ref{Standard} and \ref{MoChess}.
\begin{Thm}\label{MoChess}
Let $\mathfrak{A}$ be a vector space  with antilinear involution $J$.
Let $\omega$ be a multilinear functional on $\mathfrak{A}^{2M+1}$ with
 $M$ even. Assume the following:
\begin{itemize}
\item[{\rm (i)}] $\omega(A_1, \dots, A_{2M+1})=\omega(A_2, \dots,
	     A_{2M+1}, A_1)$. 
\item[{\rm (ii)}] There exist real linear maps $T_{+}$ and $T_-$ from
	     $\mathfrak{A}$ to $\mathfrak{A}$ such that
\begin{itemize}
\item[{\rm (a)}] For each  $A$ in $\mathfrak{A}$, 
\begin{align}
T_{\alpha}(T_{\beta}(A))=T_{\beta}(A),\ \ \alpha, \beta=+, -.
\end{align} 
\item[{\rm (b)}]
$
\omega\big(A_1, \dots, A_M, T_{\pm}(A_{M+1}), JA_M, \dots, JA_1\big)\ge 0
$.
\item[{\rm (c)}]
\begin{align}
&\Big|
\omega(A_1, \dots, A_{2M+1})
\Big|\no
\le &\omega\big(A_1, \dots, A_M, T_+(A_{M+1}), JA_M, \dots, JA_1\big)^{1/2}
\no
&\times \omega\big(JA_{2M+1}, \dots, JA_{M+2}, T_-(A_{M+1}), A_{M+2}, \dots, A_{2M+1}\big)^{1/2}.\label{ModSch}
\end{align} 
\end{itemize} 
\end{itemize} 
For each  $A_1, \dots, A_{2M+1}\in \mathfrak{A}$, we
 have\footnote{
By (a) and (b), we have $
\omega(JA, A, \dots, JA, A, T_+(JA)) \ge 0
$
 for all $A\in \mathfrak{A}$.
} 
\begin{align}
\Big|
\omega(A_1, \dots, A_{2M+1})
\Big|&\le \prod_{j=1}^{2M+1}
\omega\big(
JA_j, A_j, \dots, JA_j, A_j, T_+(JA_j)
\big)^{1/2M+1} \label{Chess1}
\end{align} 
and 
\begin{align}
\omega\big(
JA_j, A_j, \dots, JA_j, A_j, T_+(JA_j)
\big)=
\omega\big(
JA_j, A_j, \dots, JA_j,  A_j, T_-(A_j)
\big) \label{Chess2}
\end{align} 
for all $j=1, \dots, 2M+1$.
\end{Thm} 
{\it Proof.}
First, we present the proof of Eq.  (\ref{Chess1}). Without loss of generality, we  assume that 
\begin{align}
\omega\big(
JA_j, A_j, \dots, JA_j, A_j, T_+(JA_j)
\big)=1
\end{align}
  for all $j=1, \dots, 2M+1$.
We set
\begin{align*}
JA_j=A_{j+2M+1},\ \ 
&T_+(A_j)=A_{j+4M+2},\ \
T_-(A_j)=A_{j+6M+3},
\\ 
&T_+(JA_j)=A_{j+8M+4},\ \
T_-(JA_j)=A_{j+10M+5},\  j=1, \dots, 2M+1. 
\end{align*} 
A configuration $c$ is a function on $\{1, \dots, 2M+1\}$ with value in
$\{1, \dots, 12M+6\}$.

Let $z=\max_c 
|
\omega(A_{c(1)},\dots, A_{c(2M+1)})
|
$,
and let
$\tilde{c}$ be a configuration  such that 
$
z=
\big|
\omega(A_{\tilde{c}(1)},\dots, A_{\tilde{c}(2M+1)})
\big|
$. 
It suffices to show that $z=1$. 
It is easy to see  that $1\le z$.  We  show $z\le 1$.
Toward this end, set $\tilde{c}(1)=j$. Then 
we have 
\begin{align}
z=&\Big|
\omega\big(
A_{\tilde{c}(1)}, \dots, A_{\tilde{c}(2M+1)}
\big)
\Big|\no
\le& \omega\big(
A_j, A_{\tilde{c}(2)},\dots, A_{\tilde{c}(M)}, T_+(A_{\tilde{c}(M+1)}),
 JA_{\tilde{c}(M)}, \dots, JA_j
\big)^{1/2}\no
&\times \omega
\big(JA_{\tilde{c}(2M+1)},\dots, JA_{\tilde{c}(M+2)},
T_-(A_{\tilde{c}(M+1)}), A_{\tilde{c}(M+2)}, \dots, A_{\tilde{c}(2M+1)}
\big)^{1/2}\no
\le & z^{1/2} \omega
\big(
\underbrace{JA_j, A_j, A_{\tilde{c}(2)},\dots, A_{\tilde{c}(M-1)}}_{M},
 A_{\tilde{c}(M)}, \underbrace{T_+(A_{\tilde{c}(M+1)}), \dots, JA_{\tilde{c}(2)})}_{M}
\big)^{1/2}\no
\le &
z^{3/4} \omega
\big(
\underbrace{JA_j, A_j, JA_j, A_j, A_{\tilde{c}(2)}, \dots,
 }_{M}A_{\tilde{c}(M-2)},
\underbrace{A_{\tilde{c}(M-1)},
 T_+(A_{\tilde{c}(M)}), \dots JA_{\tilde{c}(2)} }_M
\big)^{1/4}\no
&\vdots\no
\le &
z^{1-2^{-(m-2)}} \omega(\underbrace{JA_j, A_j,\dots, JA_j, A_j}_{M},
 A_{\tilde{c}(2)}, 
\underbrace{*,
 \dots, *}_M)^{2^{-(m-2)}}
\no
\le & z^{1-2^{-(m-1)}}\omega(
\underbrace{JA_j, A_j,\dots, JA_j, A_j}_{M},
 T_+(A_{\tilde{c}(2)}), JA_j, A_j, \dots, JA_j, A_j 
)^{2^{-(m-1)}}\no
\le & z^{1-2^{-m}} \omega(JA_j, A_j, \dots, JA_j, A_j, T_+(JA_j), JA_j,
 A_j, \dots, JA_j, A_j)^{2^{-m}}\no
=&z^{1-2^{-m}}.
\end{align} 
Thus, we conclude that  $z\le 1$,  which implies Eq.  (\ref{Chess1}).

To show Eq. (\ref{Chess2}), we observe that 
\begin{align}
z&= \omega\big(JA_j, A_j, \dots, JA_j,  A_j, T_+(JA_j)\big)\no
&= \omega\big(\underbrace{T_+(JA_j), JA_j, A_j, \dots, JA_j}_{M}, A_j, \underbrace{JA_j, A_j, \dots, JA_j,
 A_j}_{M}\big) \no
&\le z^{1/2} \omega\big(JA_j, A_j, \dots, JA_j, A_j, T_-(A_j), JA_j,  A_j,
 \dots, JA_j, A_j\big)^{1/2}\no
&=z^{1/2} \omega\big(JA_j, A_j, \dots, JA_j, A_j, T_-(A_j)\big)^{1/2}.
\end{align} 
Thus, we conclude Eq. (\ref{Chess2}).
 $\Box$

\subsection{Proof of Theorem \ref{MainRP}: Step 1}
We  now are ready to  present the proof of   Theorem \ref{MainRP}.
The proof is divided  into three steps. 

\begin{define}{\rm 
 Let  $\partial \gamma$  be the
 contour  associated with  $\gamma\in\mathscr{S}_{\Lambda}$.
\begin{itemize}
\item $\partial \gamma_{h}=
\big\{\la \bi, \bj\ra\in \partial \gamma\, |\, 
\mbox{$\exists a\in \BbbZ$ s.t. }\bi-\bj=(a, 0)
\big\}$.
\item $\partial \gamma_v=\partial \gamma\backslash \partial \gamma_h
=\big\{\la \bi, \bj\ra\in \partial \gamma\, |\, 
\mbox{$\exists b\in \BbbZ$ s.t. }\bi-\bj=(0, b)
\big\}.
$
\item For each $\la \bi, \bj\ra\in \partial \gamma_h$, $\bi\wedge\bj$ denotes
      the site with smaller $1$-coordinate.  
      \item
      For each  $\la \bi, \bj\ra\in \partial \gamma_v$, $\bi\wedge\bj$ denotes
      the site with smaller $2$-coordinate.  

\item For each $ \alpha=h, v$, we set
\begin{align}
\partial \gamma_{\alpha, e}=\big\{\la \bi, \bj\ra\in \partial
      \gamma_{\alpha}\, |\, \bi\wedge \bj: \mathrm{even}\big\},
\ \  \partial \gamma_{\alpha, o}=\big\{\la \bi, \bj\ra\in \partial
      \gamma_{\alpha}\, |\, \bi\wedge \bj: \mathrm{odd}\big\}.
\end{align} 
Note  that $\partial \gamma= \partial \gamma_h\cup \partial \gamma_v
= (\partial \gamma_{h,e}\cup \partial \gamma_{h, o})\cup
(\partial \gamma_{v,e}\cup \partial \gamma_{v, o})
$. $\diamondsuit$
\end{itemize} 
}
\end{define} 

\begin{lemm}
To prove Theorem \ref{MainRP}, it suffices to show 
\begin{align}
\Bigg\la \prod_{\la \bi, \bj\ra\in \partial \gamma_{\alpha, \beta}} P_{\bi}^{(+)}
 P_{\bj}^{(-)}
\Bigg\ra_{\Lambda}\le \mathcal{P}_{\Lambda}^{2|\partial \gamma_{\alpha, \beta}|/|\Lambda|}
\label{ChessRed}
\end{align} 
for all $\alpha=h, v$ and $\beta=e, o$.
\end{lemm} 
{\it Proof.} By the Schwartz inequality, we have
\begin{align}
\Bigg\la \prod_{\la \bi, \bj\ra\in \partial \gamma} P_{\bi}^{(+)}
 P_{\bj}^{(-)}
\Bigg\ra_{\Lambda}
&\le
 \Bigg\la \prod_{\la \bi, \bj\ra\in \partial \gamma_{h}} P_{\bi}^{(+)}
 P_{\bj}^{(-)}
\Bigg\ra_{\Lambda}^{1/2}
\Bigg\la \prod_{\la \bi, \bj\ra\in \partial \gamma_{v}} P_{\bi}^{(+)}
 P_{\bj}^{(-)}
\Bigg\ra_{\Lambda}^{1/2}\no
&\le 
\prod_{\alpha=h, v} \prod_{\beta=e, o}
\Bigg\la \prod_{\la \bi, \bj\ra\in \partial \gamma_{\alpha, \beta}} P_{\bi}^{(+)}
 P_{\bj}^{(-)}
\Bigg\ra_{\Lambda}^{1/4}.
\end{align} 
This finishes the proof. $\Box$

\subsection{Proof of Theorem \ref{MainRP}: Step 2}

The Hilbert space for a single electron is $\ell^2(\Lambda) \oplus
\ell^2(\Lambda)$. Note  the identification 
\begin{align}
\ell^2(\Lambda) \oplus \ell^2(\Lambda)=\bigoplus_{\bj\in \Lambda}
 (\BbbC\oplus \BbbC). \label{Smallell}
\end{align} 
Recall the well-known property of  fermion Fock space:
\begin{align}
\Fock(\mathfrak{h}_1\oplus \mathfrak{h}_2)=\Fock(\mathfrak{h}_1)\otimes
 \Fock(\mathfrak{h}_2). \label{Factorize}
\end{align} 
By Eqs. (\ref{Smallell}) and (\ref{Factorize}), the fermion  Fock space can be identified as 
\begin{align}
\mathfrak{H}
=\bigotimes_{\bj\in \Lambda} \mathfrak{H}_{\bj}, \label{FctF}
\end{align} 
where $\mathfrak{H}_{\bj}=\Fock(\BbbC\oplus \BbbC)$.
Therefore, 
$\mathfrak{H}$ can be expressed as 
\begin{align}
\mathfrak{H}=\bigotimes_{k=-L}^{L-1} \mathfrak{H}_h(k), \label{Tensor}
\end{align} 
where
$\displaystyle 
 \mathfrak{H}_h(k)
=\bigotimes_{{\bj=(j_1, j_2)\in \Lambda}\atop{j_2=k}}
 \mathfrak{H}_{\bj}.
$
Let $\tilde{\mathfrak{H}}_h=\mathfrak{H}_h(k=0)$.
For each $A\in \mathfrak{B}(\tilde{\mathfrak{H}}_h)$, 
we define a linear operator on $\mathfrak{H}$  by
\begin{align}
\tau_{j}(A)=\overbrace{\underbrace{\one \otimes \cdots \otimes \one}_{L+j+1} \otimes
 A\otimes\one \otimes  \cdots \otimes  \one}^{2L},\ \  j=-L, \dots, L-1. 
\end{align}  
Here, the tensor products correspond to Eq.  (\ref{Tensor}).
We want to apply Theorem \ref{Standard} with 
\begin{align}
\mathfrak{A}=\mathfrak{B}(\tilde{\mathfrak{H}}_h),\ \ \ \ 
\omega(A_{-L}, \dots, A_{L-1})=\Bigg\la 
\prod_{j=-L}^{L-1}
\tau_{j}(A_j)
\Bigg\ra_{\Lambda}.
\end{align} 
To this end, we have to choose a suitable $J$.
Let $\xi_h$ be an antiunitary operator on $\tilde{\mathfrak{H}}_h$ defined by
\begin{align}
\xi_h\Omega_h=\Omega_h,\ \ \ \xi_h \tilde{c}_{\bj\sigma} \xi_h^{-1}=\tilde{c}_{\bj\sigma}(-1)^{N_h},
\end{align} 
where 
 $\tilde{c}_{\bj\sigma}$ is the annihilation operator on
 $\tilde{\mathfrak{H}}_h$,  $\Omega_h$ is the Fock vacuum in
 $\tilde{\mathfrak{H}}_h$, and 
$N_h=\sum_{\bj\ \mathrm{s.t.} \  j_2=0} \tilde{n
}_{\bj}$ with $\tilde{n}_{\bj}=\tilde{c}_{\bj\sigma} \tilde{c}_{\bj\sigma}$.
Now, $J$ is defined by 
$JA=\xi_h A\xi_h^{-1}$ for all  $A\in
 \mathfrak{B}(\tilde{\mathfrak{H}}_h)$.
It is easy to check that $J\tilde{n}_{\bj\sigma}=\tilde{n}_{\bj\sigma}$.

Note the following relationship: Let $\bj=(j_1, j_2)\in \Lambda^U$ such
that $j_2=k>0$. Then
\begin{align}
\vartheta_h^{-1} 
\underbrace{
n_{\bj\sigma}
}_{\in \mathfrak{B}(\mathfrak{H}^U)}\vartheta_h
=
\underbrace{ \one  \otimes \cdots \otimes \one \otimes
 J \tilde{n}_{\bj\sigma}
\otimes\overbrace{ \one \otimes \cdots \otimes \one}^{k-L-1} 
}_{\in \mathfrak{B}(\mathfrak{H}^L)}, \label{RPrelation1}
\end{align} 
where $\vartheta_h$ is defined  by Eq.  (\ref{thetah}).

\begin{Prop}\label{RPH}
Let $A_{-L}, A_{-L+1}, \dots, A_{L-1}\in \mathfrak{B}(\tilde{\mathfrak{H}}_h)$. We have 
\begin{align}
\Bigg|
\Bigg\la 
\prod_{j=-L}^{L-1}
\tau_{j}(A_j)
\Bigg\ra_{\Lambda}
\Bigg| \le \prod_{j=-L}^{L-1}
\Bigg\la
\tau_{-L}(JA_j) \tau_{-L+1}(A_j)\cdots \tau_{L-2}(JA_j) \tau_{L-1}(A_j)
\Bigg\ra_{\Lambda}^{1/2L}.
\end{align} 
\end{Prop} 
{\it Proof.}
Assumption (i) in Theorem \ref{Standard} is fulfilled  by the
translational invariance of the model.
Assumptions (ii) and (iii) in Theorem \ref{Standard} are satisfied by Theorem
\ref{RPInq2}. $\Box$
\medskip\\

For each $\ell=-L, \dots, L-1$, we define  
\begin{align}
\partial \gamma_{h, e}(\ell)
=\big\{\la \bi, \bj\ra\in \partial \gamma_{h, e}\, |\, i_2=j_2=\ell
\big\}.
\end{align} 
Trivially, we have $
\partial \gamma_{h, e}=\bigcup_{\ell=-L}^{L-1} \partial \gamma_{h, e}(\ell).
$

\begin{Prop}\label{MoChess2}
We have 
\begin{align}
\Bigg\la
\prod_{\la \bi, \bj\ra\in \partial \gamma_{h, e}} P_{\bi}^{(+)} P_{\bj}^{(-)}
\Bigg\ra_{\Lambda}
\le \prod_{\ell\, :\,  \partial \gamma_{h, e}(\ell)\neq \emptyset}
\Bigg\la 
\prod_{\la \bi, \bj\ra\in \partial \gamma_{h,
 e}(\ell)}\prod_{k=-L}^{L-1} 
P_{(i_1, k)}^{(+)} P_{(j_1, k)}^{(-)}
\Bigg\ra_{\Lambda}^{1/2L}. \label{DD}
\end{align} 
Here, $i_1$ and $j_1$ on the right-hand side of Eq. (\ref{DD}) are related
 to $\bi$ and $\bj$ by $\bi=(i_1, \ell)$ and $\bj=(j_1, \ell)$.
\end{Prop} 
{\it Proof.}
Let $c_{\sigma}\ (c^*_{\sigma})$ be the annihilation(creation) operator
 in $\Fock(\BbbC
\oplus \BbbC)$. The number operator is $n=\sum_{\sigma=\uparrow,
 \downarrow} n_{\sigma}
$ with $n_{\sigma}=c_{\sigma}^* c_{\sigma}$. Let $q=n-\one$.
Corresponding to Eq.  (\ref{FctF}), we have $q_{\bj}=\bigotimes_{\bi\in
 \Lambda} q^{\delta_{\bi\bj}}
$, where $q^{\delta_{\bi\bj}}=q$ if $\bi=\bj$,
 $q^{\delta_{\bi\bj}}=\one$ if $\bi\neq \bj$.
Let $E_q(\cdot)$ be the spectral measure of $q$. We set
$
P^{(+)}=E_q(\{0, +1\}),\ P^{(-)}=E_q(\{-1\})
$. Trivially, we have $P_{\bj}^{(\omega)}=\bigotimes_{\bi\in \Lambda}
(P^{(\omega)})^{\delta_{\bi\bj}}
$.

Let $\Lambda_h=\{-L, -L+1, \dots, L-1\}$. Let $A$ be a linear operator
in $\Fock(\BbbC\oplus \BbbC)$.
For each $\mathscr{I}\subseteq \Lambda_h$, we define $
[A]_k^{\mathscr{I}} \in \mathfrak{B}(
\Fock(\BbbC\oplus \BbbC)
)
$ by 
\begin{align}
[A]_k^{\mathscr{I}}=
\begin{cases}
A    &\mbox{if $k\in \mathscr{I}$}\\
\one &\mbox{if $k\notin \mathscr{I}$}
\end{cases}. \label{AKI}
\end{align} 
In this proof, an operator of the form
$
\bigotimes_{k\in \Lambda_h} [A]_k^{\mathscr{I}} [B]_k^{\mathscr{I}'}
$
will  play an important role.

Set  
\begin{align} 
B_{\ell}=
\begin{cases}\displaystyle 
\prod_{\la \bi, \bj\ra\in \partial \gamma_{h, e}(\ell)}P_{\bi}^{(+)}
 P_{\bj}^{(-)} & \mbox{if $\partial \gamma_{h, e}(\ell)\neq \emptyset$}\\
\displaystyle\ \ \ \  \one &  \mbox{if $\partial \gamma_{h, e}(\ell)= \emptyset$}.
\end{cases} 
\end{align} 
Note that 
\begin{align}
\prod_{\la \bi, \bj\ra\in \partial \gamma_{h, e}} P_{\bi}^{(+)}
 P_{\bj}^{(-)}
=\prod_{\ell=-L}^{L-1} B_{\ell}.
\end{align} 
We  rewrite  $B_{\ell }$ by using the notation of Eq.  (\ref{AKI}),
because  the new
expression is convenient for our proof.
To this end, write $\partial \gamma_{h, e}(\ell)=\{\la \bi^{(1)},
\bj^{(1)}\ra, \dots, \la \bi^{(m)},\bj^{(m)}\ra
\}$. We also  have $\bi^{(\alpha)}=(i_1^{(\alpha)}, \ell)$ and
$\bj^{(\alpha)} =(j_1^{(\alpha)}, \ell)$ for each $\alpha=1, \dots, m$.
Let $\mathscr{O}_+^{(\ell)}=\{i_1^{(1)}, \dots, i_1^{(m)}\},
\ \mathscr{O}_-^{(\ell)}=\{j_1^{(1)}, \dots, j_1^{(m)}\}$.
Now, let
\begin{align}
A_{\ell}=\bigotimes_{k\in \Lambda_h}
 \Big[P^{(+)}\Big]_k^{\mathscr{O}_+^{(\ell)}}\Big[P^{(-)}\Big]_k^{\mathscr{O}_-^{(\ell)}}
 \in \mathfrak{B}(\tilde{\mathfrak{H}}_h).
\end{align} 
If $\partial \gamma_{h, e}(\ell)=\emptyset$, we simply set $A_{\ell}=\one$.
$B_{\ell}$ can be expressed as $B_{\ell}=\tau_{\ell}(A_{\ell})$.
Note  that 
   $JA_{\ell}=A_{\ell}$.
Now,  we  apply Proposition \ref{RPH}. Because  
\begin{align}
\Bigg\la
\prod_{k=-L}^{L-1} \tau_{k}(A_{\ell})
\Bigg\ra_{\Lambda}
=
\begin{cases}
\displaystyle 
\Bigg\la
\prod_{k=-L}^{L-1} 
\prod_{\la \bi, \bj\ra\in \partial \gamma_{h,
 e}(\ell)}
P_{(i_1, k)}^{(+)} P_{(j_1, k)}^{(-)}\Bigg\ra_{\Lambda}
& \mbox{if $\partial \gamma_{h,
e}(\ell)\neq \emptyset$  }\\
\displaystyle \ \ \ \ \  1 & \mbox{if $\partial \gamma_{h,
e}(\ell)=\emptyset$  }\\
\end{cases}, 
\end{align} 
we obtain the assertion in the proposition. $\Box$

\subsection{Proof of Theorem \ref{MainRP}: Step 3}
In this step, we identify $\mathfrak{H}$ as 
\begin{align}
\mathfrak{H}=\bigotimes_{k=1}^L \mathfrak{H}_v(k), \label{Verticle}
\end{align}
where 
$\displaystyle 
 \mathfrak{H}_v(k) 
=\bigotimes_{{\bj=(j_1, j_2)\in \Lambda}
\atop{j_1=-L+2k-2,\ -L+2k-1}} \mathfrak{H}_{\bj}.
$
  Recall that $L=2M+1$. 
We suppose that $M$ is even. 
Set $\hat{\mathfrak{H}}_v=\mathfrak{H}_v(k=\frac{L+1}{2})$.
Note  that $
\hat{\mathfrak{H}}_v=\h_L\otimes \h_R
$, where $\h_L
=\bigotimes_{\bj\in \Lambda,\  j_1=-1} \mathfrak{H}_{\bj}
$ and 
$\h_R
=\bigotimes_{\bj\in \Lambda,\  j_1=0} \mathfrak{H}_{\bj}
$ .

For each $A\in \mathfrak{B}(\hat{\mathfrak{H}}_v)$, we define a linear operator  on
$\mathfrak{H}$ by 
\begin{align}
\eta_j(A):=
\overbrace{
\underbrace{\one \otimes \cdots \otimes \one}_{j-1} \otimes A\otimes \one
 \otimes \cdots \otimes \one}^{2M+1} ,\ \ j=1, \dots,
 2M+1.\label{Aotimes}
\end{align} 
Here, the tensor products in Eq.  (\ref{Aotimes}) correspond to Eq.  (\ref{Verticle}).
Let $\xi_v$ be an antiunitary transformation from  $\h_L$ onto $\h_R$
that is  defined
by 
\begin{align}
\xi_v \hat{\Omega}_L=\hat{\Omega}_R,\ \ \ \xi_v^{-1} \hat{c}_{\bj \sigma} \xi_v
 =\hat{c}_{r_v(\bj) \sigma}(-1)^{\hat{N}_L},\ \ \bj\in \Lambda\
 \mbox{s.t. $j_1=0$},
\label{Defrv} 
\end{align} 
where  $\hat{c}_{\bj\sigma}$ is the annihilation operator on
$\hat{\mathfrak{H}}_v$,  $\hat{\Omega}_L$ ($\hat{\Omega}_R$) is the Fock vacuum in $
\mathfrak{h}_L$ ($\mathfrak{h}_R$),  and 
$\hat{N}_L=\sum_{\sigma}\sum_{\bj\ \mathrm{s.t.}\ j_1=-1} \hat{n}_{\bj\sigma}$  with
$\hat{n}_{\bj\sigma}=\hat{c}_{\bj\sigma}^*\hat{c}_{\bj\sigma}
$.
Here, $r_v$ is defined by Eq. (\ref{thetav})\footnote{
To be precise, $r_v$ in Eq. (\ref{Defrv}) is a restriction of $r_v$ to 
$\{\bj=(j_1, j_2)\in \Lambda\, | \, j_1=0\}$.
}.

For each $A\in \mathfrak{B}(\h_L)$ and $B\in \mathfrak{B}(\h_R)$, we set
\begin{align}
T_+(A\otimes B)=A\otimes (\xi_v A\xi_v^{-1}),\ \ \ T_-(A\otimes
 B)=(\xi_v^{-1} B \xi_v)\otimes B.
\end{align} 
$T_{\pm}$ are real linear maps on $\mathfrak{B}(\hat{\mathfrak{H}}_v)$
that satisfy $T_{\alpha} \circ T_{\beta}=T_{\beta}$.
We define an antilinear involution $J$ on
$\mathfrak{B}(\hat{\mathfrak{H}}_v)$ by 
\begin{align}
J(A\otimes B)=(\xi_v^{-1} B\xi_v)\otimes (\xi_v A \xi_v^{-1}).
\end{align} 

Let $\bj\in \Lambda$ such that $j_1=-1$. We have
\begin{align}
n_{\bj}\vartheta_v n_{\bj} \vartheta_v^{-1}
=
\underbrace{
(\one\otimes \cdots  \otimes \one \otimes  \hat{n}_{\bj})
}_{\in \mathfrak{B}(\mathfrak{H}_L)}
\otimes
\underbrace{ ((\xi_v \hat{n}_{\bj} \xi_v^{-1})\otimes \one \otimes \cdots
 \otimes \one)}_{\in \mathfrak{B}(\mathfrak{H}_R)};
\end{align} 
let $\la \bi, \bj\ra\in \gamma_{h,e}$ such that $\bi\wedge \bj=i_1=-2$.
 We have
\begin{align}
\vartheta_v
n_{\bi } n_{\bj}
\vartheta_v^{-1}
= 
\underbrace{
\overbrace{
(\xi_v^{-1} \hat{n}_{\bi} \xi_v)\otimes (\xi_v \hat{n}_{\bj}
 \xi_v^{-1})
}^{\in \mathfrak{B}(\hat{\mathfrak{H}}_v)}
\otimes \one \otimes \cdots \otimes \one 
}_{\in \mathfrak{B}(\mathfrak{H}_R)},
\end{align} 
where $\vartheta_v$ is defined by Eq. (\ref{thetav}).

\begin{Prop}\label{RPV}
Let  $A_1, \dots, A_{2M+1} \in \mathfrak{B}(\hat{\mathfrak{H}}_v)$. We have 
\begin{align}
&\Bigg|
\Bigg\la 
\prod_{\alpha=1}^{2M+1} \eta_{\alpha}(A_{\alpha})
\Bigg\ra_{\Lambda}
\Bigg|\no
\le &\prod_{\alpha=1}^{2M+1} \Bigg\la
\eta_1(JA_{\alpha}) \eta_2(A_{\alpha})\cdots \eta_{2M-1}(JA_{\alpha})\eta_{2M}(A_{\alpha})
\eta_{2M+1}\big(T_+(JA_{\alpha})\big)
\Bigg\ra_{\Lambda}^{1/2M+1}.
\end{align} 
\end{Prop} 
{\it Proof.} We apply Theorem \ref{MoChess} with 
\begin{align}
\mathfrak{A}=\mathfrak{B}(\hat{\mathfrak{H}}_v),\ \ \ \omega(A_1, \dots A_{2M+1})
=\Bigg\la \prod_{\alpha=1}^{2M+1} \eta_{\alpha}(A_{\alpha}) \Bigg\ra_{\Lambda}.
\end{align} 
Note  that the assumptions (b) and (c) in Theorem \ref{MoChess} are  satisfied by Theorem \ref{RPInq}. $\Box$
\medskip\\

For each $\alpha=1, \dots, 2M+1$, we set
\begin{align}
\partial \gamma_{h, e}(\ell; \alpha)
=\big\{
\la \bi, \bj\ra\in \partial \gamma_{h, e}(\ell)\, |\, \bi\wedge \bj =-L-2+2\alpha
\big\}.
\end{align} 
Note  that\footnote{
$\# S$ is the cardinality of  set $S$.
} $\#\partial \gamma_{h, e}(\ell; \alpha) =0$ or $1$.
Let us define a linear operator  $C_{\alpha}$ by 
\begin{align}
C_{\alpha}=
\begin{cases}
\displaystyle 
\prod_{k=-L}^{L-1} P_{(i_1, k)}^{(+)}P_{(j_1, k)}^{(-)}&
 \mbox{if $\partial
 \gamma_{h, e}(\ell; \alpha)\neq \emptyset$ }\\
\ \ \ \ \one &
 \mbox{if $\partial
 \gamma_{h, e}(\ell; \alpha)= \emptyset$ }
\end{cases}, 
\end{align} 
where  
 $i_1, j_1$ satisfy $\min\{i_1, j_1\}=-L-2+2\alpha$
and $\partial \gamma_{h, e}(\ell; \alpha)=\big\{\big\la (i_1, \ell), (j_1, \ell)\big\ra\big\}$.
Note  that
 \begin{align}
\prod_{\la \bi, \bj\ra\in \partial \gamma_{h,
 e}(\ell)}\prod_{k=-L}^{L-1} 
P_{(i_1, k)}^{(+)} P_{(j_1, k)}^{(-)}
=\prod_{\alpha=1}^{2M+1}C_{\alpha}.
\end{align} 

We consider the case where $\partial \gamma_{h, e}(\ell;
\alpha) \neq \emptyset$.
Write $\partial \gamma_{h, e}(\ell; \alpha)=\{\la\bi^{[\alpha]}, \bj^{[\alpha]}\ra\}$.
Suppose first that $j_1^{[\alpha]}>i_1^{[\alpha]}=-L-2+2\alpha$.
Let 
\begin{align}
\mathscr{P}^{(+)}=\bigotimes_{k=-L}^{L-1} P^{(+)},\ \ \mathscr{P}^{(-)}
 =\bigotimes_{k=-L}^{L-1} P^{(-)},
\end{align} 
where $P^{(\pm)}$ is  defined in the proof of Proposition \ref{MoChess2}.
Let us define a linear operator $A\in \mathfrak{B}(\hat{\mathfrak{H}}_v)$ by
$A=\mathscr{P}^{(+)}\otimes \mathscr{P}^{(-)}$.
Here, we regard $\mathscr{P}^{(+)}$ ($\mathscr{P}^{(-)}$)
as a linear operator on $\mathfrak{B}(\mathfrak{h}_L)$
($\mathfrak{B}(\h_R)$). Evidently, 
$C_{\alpha}$ can be expressed as
$C_{\alpha}=\eta_{\alpha}(A)$.
 Because  $JA=\mathscr{P}^{(-)}\otimes \mathscr{P}^{(+)}$ and 
$T_+(JA)=\mathscr{P}^{(-)}\otimes \mathscr{P}^{(-)}$,
we have
\begin{align}
\Bigg\la
\eta_1(JA) \eta_2(A)\cdots \eta_{2M-1}(JA)\eta_{2M}(A)
\eta_{2M+1}(T_+\big(JA)\big)
\Bigg\ra_{\Lambda}
=\Big\la \mathbf{P}_{\Lambda}^{(-)}\Big\ra_{\Lambda}. \label{Eta1}
\end{align} 

Conversely, if $i_1^{[\alpha]}>j_1^{[\alpha]}=-L-2+2\alpha$, 
 we see that $C_{\alpha}$ can be expressed as $C_{\alpha}=\eta_{\alpha}(B)$
with 
$B=\mathscr{P}^{(-)}\otimes \mathscr{P}^{(+)}$. Moreover, 
\begin{align}
\Bigg\la
\eta_1(JB) \eta_2(B)\cdots \eta_{2M-1}(JB)\eta_{2M}(B)
\eta_{2M+1}(T_+\big(JB)\big)
\Bigg\ra_{\Lambda}
=\Big\la \mathbf{P}_{\Lambda}^{(+)}\Big\ra_{\Lambda}. \label{Eta2}
\end{align}

We apply Proposition  \ref{RPV} with
 $A_{\alpha}=A$ if $j_1^{[\alpha]}>i_1^{[\alpha]}$, $A_{\alpha}=B$ if $j_1^{[\alpha]} <i_1^{[\alpha]}$.
If  $\partial \gamma_{h, e}(\ell)\neq \emptyset$, then at least
 one $\alpha$ exists such that $
\partial \gamma_{h, e}(\ell; \alpha)\neq \emptyset
$. Thus, by Eqs. (\ref{Eta1}),  (\ref{Eta2}) and Proposition \ref{RPV},
we have
\begin{align}
\Bigg\la\prod_{\la \bi, \bj\ra\in \partial \gamma_{h,
 e}(\ell)}\prod_{k=-L}^{L-1} 
P_{(i_1, k)}^{(+)} P_{(j_1, k)}^{(-)}\Bigg\ra_{\Lambda}
\le  \mathcal{P}_{\Lambda}^{1/2M+1}.
\end{align} 
Thus, by Proposition \ref{MoChess2}, we obtain 
\begin{align}
\Bigg\la
\prod_{\la \bi, \bj\ra\in \partial \gamma_{h, e}} P_{\bi}^{(+)} P_{\bj}^{(-)}
\Bigg\ra_{\Lambda}
\le \mathcal{P}_{\Lambda}^{2|\partial \gamma_{h, e}|/|\Lambda|}.
\end{align} 
This finishes the proof of Eq.  (\ref{ChessRed}) for  the case where
$\alpha=h$ and $\beta=e$.
In a similar way, Eq. (\ref{ChessRed}) can be proven  in the remaining three
cases. $\Box$

\section{Proof of Theorem \ref{LocEstRep}}\label{PfLocal}
\setcounter{equation}{0}

 We apply the  {\it  principle of exponential localization} that was   established
  by Fr\"ohlich and Lieb \cite{FL}, which is stated as follows.
\begin{Thm}\label{Principle}
Let $A$ and $B$ be self-adjoint operators on a Hilbert space
 $\mathscr{H}$
such that 
\begin{itemize}
\item[{\rm (i)}] $A\ge 0$,
\item[{\rm (ii)}] $\pm B\le \vepsilon A$ with $\vepsilon\in [0, 1)$.
\end{itemize} 
Suppose that 
\begin{align}
(A+B)\psi =\lambda \psi,\ \ \|\psi\|=1.
\end{align}  
Choose some $\rho>\lambda $ such that
 $\gamma:=\vepsilon\rho(\rho-\lambda)^{-1}<1$.
Let $P_{\rho}=E_A[\rho, \infty)$, the spectral measure of $A$
 corresponding to $[\rho, \infty)$, and let $\mathcal{M}_{\rho}=\ran
 P_{\rho}$.  Finally, let $\mathcal{N}$ be the closed subspace 
such that 
\begin{itemize}
\item[{\rm (iii)}] 
$\{B(A-\lambda)^{-1}\}^j \mathcal{N} \subseteq \mathcal{M}_{\rho}, \
	     j=1, \dots, d-1$ with $d\ge 1$.\footnote{
More precisely, for all $\phi\in \mathcal{N}$,
	     $\{B(A-\lambda)^{-1}\}^{-j}\phi \in \mathcal{M}_{\rho}$
 for $j=1, \dots, d$.}
\end{itemize} 
Then $\la \psi|P_{\mathcal{N}}\psi\ra\le \gamma^d$, where
 $P_{\mathcal{N}}$
 is the orthogonal projection onto $\mathcal{N}$.
\end{Thm}

 Theorem \ref{LocEstRep} can be proven  by applying Theorem \ref{Principle}.
We begin with the following lemma:

\begin{lemm}\label{WGS}
Denote by $\underline{e}^W$ the lowest eigenvalue for $W$.
We obtain  the following:
\begin{itemize}
\item[{\rm (i)}]
$\displaystyle 
\underline{e}^W
=
-\Big(S+\frac{\Delta}{2}\Big)|\Lambda|
$.
\item[{\rm (ii)}]
The Fock vacuum $\Omega$ is  the ground state of $W$: $W\Omega=\underline{e}^W\Omega$.
 
\end{itemize} 
\end{lemm} 
{\it Proof.} Use Eq.  (\ref{Wex}). $\Box$
\medskip\\

We choose  $A$ and $B$  as 
\begin{align}
A=W
-\underline{e}(t=1),\ \ \ \ 
B=T=(-t)\sum_{\la \bi,\bj\ra} \sum_{\sigma=\uparrow,\, \downarrow}
\big(
c_{\bi\sigma}^* c_{\bj \sigma}^*
+
c_{\bj\sigma} c_{\bi \sigma}
\big),
 \end{align} 
where $\underline{e}(t=1)$ is the lowest eigenvalue  for
$\tilde{H}$ with $t=1$.
In the remainder of this section, we present checks of  every assumption in
Theorem \ref{Principle}.

\begin{lemm}
We have the following:
\begin{itemize}
\item[{\rm (i)}] $A\ge 0$

\item[and] 
\item[{\rm (ii)}] $
\pm B\le tA
$.
\end{itemize}
\end{lemm}
{\it Proof.} (i) By Lemma \ref{WGS}, we have 
 $
\la \Omega|W\Omega\ra= \underline{e}^W
$. Because  $\la \Omega|B\Omega\ra=0$, we have
\begin{align}
\underline{e}^W=\la \Omega|W\Omega\ra=\la \Omega|\tilde{H}^{t=1}\Omega\ra
\ge \underline{e}(t=1).
\end{align} 
Thus, we obtain item  (i).

(ii) We see  that 
\begin{align}
A+t^{-1} B=\tilde{H}^{t=1}-\underline{e}(t=1) \ge 0,
\end{align} 
which implies $-B\le tA$.

Let $u=\exp\big\{i\pi \sum_{\sigma} \sum_{\bj\in \Lambda_{\rm{o}}}
n_{\bj\sigma}\big\}$. We have $
u Bu^{-1}=-B
$ and $uAu^{-1}=A$. Thus, 
\begin{align}
0\le u(A+t^{-1}B )u^{-1}=A-t^{-1}B,
\end{align} 
which implies $B\le tA$.
 $\Box$

\begin{lemm}\label{Basic2}
We have 
 $\displaystyle 
\underline{e}(t=1)\ge -8 |\Lambda|-S|\Lambda|-\frac{\Delta}{2}|\Lambda|
$.
\end{lemm}
{\it Proof.}
Because  $\| B\|\le  8t  |\Lambda|$ and
$\underline{e}^W=-(S+\Delta/2)|\Lambda|$, we obtain the result. $\Box$
\medskip\\

Let 
\begin{align}
x=\frac{1}{|\Lambda|} \{\underline{e}^W-\underline{e}(t=1)\}, \ \ \
\rho= \underline{e}^W -\underline{e}(t=1)+n \delta |\Lambda|.
\end{align}
By Lemma \ref{Basic2}, we have
$
x \le 8.
$
Thus, we obtain the following:

\begin{lemm}
For each $\lambda$ with $\lambda\le  x|\Lambda|+\delta|\Lambda|$, we define
$\displaystyle 
\gamma=t \frac{\rho}{\rho-\lambda}.
$
Then we have
\begin{align}
\gamma\le t \Bigg\{1+\frac{8+\delta}{(n-1)\delta}\Bigg\}.
\end{align}
\end{lemm}

Let
\begin{align}
\mathcal{E}^A\Big(\mb{P}_{\Lambda}^{(\omega)}\Big)=\min
 \mathrm{spec}\Big(\mb{P}_{ \Lambda}^{(\omega)} W
 \mb{P}_{\Lambda}^{(\omega)}\Big) -\underline{e}(t=1),\ \ \
 \omega=+, -.
\end{align}

\begin{lemm}\label{ErhoDiff}
We have 
$\displaystyle 
\mathcal{E}^A\Big(\mb{P}_{\Lambda}^{(\omega)}\Big)-\rho
\ge  
 \Big(\mathscr{J}-n\delta\Big)|\Lambda|
+\mathcal{O}(|\Lambda|^{1/2})
$ for each $\omega=+, -$, where $\mathscr{J}$
is defined by Eq. (\ref{DefJ}).
\end{lemm}
{\it Proof.}
Recall Eq.  (\ref{DefP}).
We can  check that 
\begin{align}
\min
 \mathrm{spec}\Big(\mb{P}_{ \Lambda}^{(\pm)} W
 \mb{P}_{\Lambda}^{(\pm)}\Big)
\ge
\begin{cases}\displaystyle 
-\Big(S+\frac{\Delta}{4}\Big)|\Lambda|+\frac{V}{4}|\Lambda|+\mathcal{O}(|\Lambda|^{1/2})
 & \mbox{if $S\ge 0$}\\
\displaystyle 
-\frac{1}{2}\Big(S+\frac{\Delta}{2}\Big)|\Lambda|+\frac{V}{4}|\Lambda|+\mathcal{O}(|\Lambda|^{1/2})
 & \mbox{if $S<0$}
\end{cases}, 
\end{align} 
which gives  the desired result. $\Box$
\medskip\\

Because  $\mathscr{J}>0$ by the assumption that  $S+\frac{\Delta}{2}>0$, we obtain the following:

\begin{coro}
Let $\mathcal{N}^{(\pm)}=\ran\Big(\mb{P}_{\Lambda}^{(\pm)}\Big)$.
 If $|\Lambda| $ is sufficiently large such
 that 
\begin{align}
 \big(\mathscr{J}-n\delta\big)|\Lambda|+\mathcal{O}(|\Lambda|^{1/2})\ge
 0,
\end{align}  then we have
$
\mathcal{N}^{(\pm)}\subseteq \mathcal{M}_{\rho}.
$

\end{coro}

\begin{Prop}\label{EShift}
Let $G=6(|S|+V)+\Delta$.
We have
$B \{\ran E_A[e, \infty) \} \subseteq \ran E_A[e-G, \infty)$. That is,  
if $\psi\in \ran E_A[e, \infty)$, then $B\psi\in \ran E_A[e-G, \infty)$.
\end{Prop} 
{\it Proof.}
For each  $\mb{m}=\{m_{\bj}\}_{\bj\in \Lambda}\in \{-1, 0,
1\}^{\Lambda}$, we set 
$
\mathfrak{H}(\mb{m})=\ran\Big[
\prod_{\bj\in \Lambda} E_{q_{\bj}}(\{m_{\bj}\})
\Big]
$. We have $\mathfrak{H}=\bigoplus_{\mb{m}\in \{-1, 0, 1\}^{\Lambda}}
\mathfrak{H}(\mb{m})
$
and 
\begin{align}
W \restriction \mathfrak{H}(\mb{m}) =e(\mb{m}),\ \ \ e(\mb{m})=
-S\sum_{\bj\in \Lambda} m_{\bj}^2
+\frac{V}{2}\sum_{\la \bi, \bj\ra}
 (m_{\bi}-m_{\bj})^2+\frac{\Delta}{2}\sum_{\bj\in \Lambda} m_{\bj},
\end{align} 
where $W \restriction \mathfrak{H}(\mb{m})$ is the restriction of $W$
onto $\mathfrak{H}(\mb{m})$.  Since  $
\mathrm{spec}(W)=\{e(\mathbf{m})
\, |\, \mathbf{m}\in \{-1, 0, 1\}^{\Lambda}
\}
$, 
 we have 
\begin{align}
\ran E_A[e, \infty)=\bigoplus_{{{\bf m}\in \{-1, 0, 1\}^{\Lambda}}\atop{
e(\mb{m})-\underline{e}(t=1) \ge e
}}\mathfrak{H}(\mb{m})
.
\end{align}

To discuss how the linear operator $B$ maps $\mathfrak{H}(\mb{m})$
for each $\mb{m}\in \{-1, 0, 1\}^{\Lambda}$,  we introduce  the
notation  $\mathcal{M}=\big\{\emptyset,
\mb{m}\ |\  \mb{m}\in
\{-1, 0, 1\}^{\Lambda}\big\}$.
 Note that the operator $B$
consists of $c_{\bi \sigma}^* c_{\bj\sigma}^*$ and $
c_{\bj\sigma} c_{\bi\sigma}
$. For each $\mb{m}\in \mathcal{M}$, there exists an $\mb{m}'\in
\mathcal{M}$
such that $c_{\bi\sigma}^{\#} c_{\bj\sigma}^{\#}\mathfrak{H}(\mb{m})\subseteq \mathfrak{H}(\mb{m}')$,
where  $\mathfrak{H}(\mathbf{m}')=\{0\}$ if $\mathbf{m}'=\emptyset$.
More precisely, $\mb{m}'$ is of the form $\mb{m}'=\{m_{\boldsymbol{k}}\pm
\delta_{{\boldsymbol k}\bi }\pm \delta_{{\boldsymbol k}\bj
}\}_{{\boldsymbol k}\in \Lambda}$, where $\delta_{{\boldsymbol k}\bi }$
is the Kronecker delta.
If $ m_{\boldsymbol{k}}\pm
\delta_{{\boldsymbol k}\bi }\pm \delta_{{\boldsymbol k}\bj
}=\pm 2$ for ${\boldsymbol k}=\bi$ or $\bj$, then we understand that
$\mb{m}'=\emptyset$. A naive  estimate tells us that 
$
|
e(\mb{m})-e(\mb{m}')
|\le 6(|S|+V)+\Delta
$. Thus, if $\psi\in \ran E_A[e,\infty)$, then $B\psi \in \ran E_A[e-G,
\infty)$. $\Box$

\begin{coro}\label{LocEst2}
Suppose that $\displaystyle
\mathscr{J}-n\delta >0$.
For sufficiently large $\Lambda$, we have
$\{B(A-\lambda)^{-1}\}^j \mathcal{N}^{(\pm)} \subseteq \mathcal{M}_{\rho}, \
	     j=1, \dots, d-1$ with $d\in \BbbN$ which satisfies
\begin{align} 
d>
\frac{1}{G} \Big(\mathscr{J}-n\delta\Big) |\Lambda|+
\mathcal{O}(|\Lambda|^{1/2})
.
\end{align} 
\end{coro} 
{\it Proof.}  
By Proposition \ref{EShift}, we remark that if $\psi\in \ran
E_A\Big[\mathcal{E}^A\Big(\mb{P}_{\Lambda}^{(\omega)}\Big), \infty\Big)$, then
$B^{\ell} \psi\in \ran E_A\Big[
\mathcal{E}^A\Big(\mb{P}_{ \Lambda}^{(\omega)}\Big)-G \ell, \infty
\Big)$ for each $\ell\in \BbbN$.
If   $d$  satisfies
\begin{align}
\mathcal{E}^A\Big(\mb{P}_{\Lambda}^{(\omega)}\Big)-Gd
< \rho \le \mathcal{E}^A\Big(
\mb{P}_{\Lambda}^{(\omega)}
\Big)-G(d-1),
 \end{align} 
then it holds that $\{B(A-\lambda)^{-1}\}^{j} \mathcal{N}^{(\pm) } \subset
 \mathcal{M}_{\rho}$ for $j=1, \dots, d-1$, where $\mathcal{N}^{(\pm)}=\ran(\mathbf{P}_{\Lambda}^{(\pm)})$.
 Thus, we have, by Lemma \ref{ErhoDiff},  
\begin{align}
d> \frac{1}{G}\bigg\{
\Big(\mathscr{J}-n\delta\Big)|\Lambda|+\mathcal{O}\Big(|\Lambda|^{1/2}\Big)
\bigg\}.
\end{align}  
This completes the proof. $\Box$

\begin{flushleft}
{\it Completion of  proof of Theorem \ref{LocEstRep} }
\end{flushleft} 
Note  that 
$\Tr_{\mathfrak{H}}[\mb{P}_{\Lambda}^{(\omega)} E_{\delta}] 
=\sum_{n} \la \psi_n| \mb{P}_{\Lambda}^{(\omega)} \psi_n\ra
$, where $\{\psi_n\}_n$ is a complete orthonormal system of $\ran E_{\delta}$.
By Theorem \ref{Principle} and Corollary \ref{LocEst2}, we have
$\la  \psi_n| \mb{P}_{\Lambda}^{(\omega)} \psi_n\ra
\le \gamma ^d
 $
for all $n$.
Therefore, it holds that 
$
\Tr_{\mathfrak{H}}[\mb{P}_{\Lambda}^{(\omega)} E_{\delta}] \le
4^{|\Lambda|} \gamma^d
$.
Finally, we choose $\delta=\beta^{-\xi}$ and $\displaystyle n=\mathscr{J}\eta \beta^{\xi}$ for arbitrary
$\xi, \eta\in (0, 1)$. $\Box$

\section{Proof of Theorem \ref{Strategy2} {\bf (B)}} \label{Proof(B)}
\setcounter{equation}{0}

We divide the proof into two cases. In this section,  we present proofs
of  the following
Theorems:
\begin{Thm}\label{QP+}
Assume that $S\ge 0$ and  $\displaystyle \frac{\Delta}{2}+S>0$.
For any $\vepsilon>0$, 
we have 
$
\Big\la P_{{\bf o}}^{(0)}\Big\ra_{\Lambda}<\vepsilon
$,
provided that $\beta$ and $t^{-1}$ are  sufficiently large.
\end{Thm} 
\begin{flushleft}
and 
\end{flushleft} 

\begin{Thm}\label{SP-}
Assume that $S<0$. Moreover, assume that $\displaystyle \frac{\Delta}{2}-|S|>0$.
For any $\vepsilon>0$,  we have 
$
\Big\la P_{{\bf o}}^{(0)}\Big\ra_{\Lambda}<\vepsilon
$,
provided that $\beta$ and $t^{-1}$  are   sufficiently large.
\end{Thm} 

\subsection{Proof of Theorem \ref{QP+}}

\begin{lemm}\label{Smallness}
Assume that $
|1-\la q_{{\bf o}}^2\ra_{\Lambda}|<\vepsilon
$.
We have $\Big\la P^{(0)}_{ \bj}\Big\ra_{\Lambda} <\vepsilon$.
\end{lemm} 
{\it Proof.} Note  that, by the spectral theorem, we have
\begin{align}
\la q_{{\bf o}}^2\ra_{\Lambda}=\Big\la P_{{\bf o}}^{\lambda=+1}\Big\ra_{\Lambda}
+\Big\la P_{{\bf o}}^{\lambda=-1}\Big\ra_{\Lambda}.
\end{align} 
Since  $
P_{{\bf o}}^{(0)}+P_{ {\bf o}}^{\lambda=+1}
+ P_{{\bf o}}^{\lambda=-1}=\one
$, we obtain the desired result. $\Box$

\begin{lemm}\label{Thermo}
We have $
\displaystyle
\la q_{{\bf o}}^2\ra_{\Lambda}^{1/2}
\ge 1-\frac{8 t}{S+\frac{\Delta}{2}}-\frac{\ln 4}{\beta (S+\frac{\Delta}{2})}
$. 
\end{lemm} 
{\it Proof.}
Because  $\displaystyle 
\frac{V}{2}\sum_{\la \bi, \bj\ra}(q_{\bi}-q_{\bj})^2\ge 0$, we have
\begin{align}
\frac{\la -\tilde{H}\ra_{\Lambda}}{|\Lambda|} \le 8 t+S\la q_{\bf o}^2\ra_{\Lambda}-\frac{\Delta}{2} \la q_{\bf o}\ra_{\Lambda}.
\end{align} 
By using the fact $|\la q_{\bf o}\ra_{\Lambda}| \le \la q_{\bf o}^2\ra_{\Lambda}^{1/2}$, we obtain
\begin{align}
\frac{\la -\tilde{H}\ra_{\Lambda}}{|\Lambda|} \le 
8
 t+\Big(S+\frac{\Delta}{2}\Big) \la q_{\bf o}^2\ra_{\Lambda}^{1/2},\label{Th1} 
\end{align} 
where we have used the fact that $x\le \sqrt{x}$ for all $x\in [0, 1]$.

Because  $\Omega$ is the ground state of $W$ and $\la
\Omega| T \Omega\ra=0$, we have $\displaystyle 
\la
\Omega|\tilde{H}\Omega\ra=-\Big(S+\frac{\Delta}{2}\Big)|\Lambda|$. Thus,
 by the Peierls--Bogoliubov inequality \cite{Simon2}, we have
\begin{align}
\Tr_{\mathfrak{H}}
\big[
e^{-\beta\tilde{H}}
\big]\ge
 e^{-\beta \la \Omega|\tilde{H}\Omega\ra}
=e^{\beta(S+\frac{\Delta}{2})}. \label{Th2}
\end{align} 
Conversely, because of the convexity of $\ln \Tr[ e^{-A}]$,
we have
\begin{align}
\la -\beta \tilde{H}\ra_{\Lambda}
\ge \ln \Tr_{\mathfrak{H}} \big[e^{-\beta
 \tilde{H}}\big]-|\Lambda |\ln 4. \label{Th3}
\end{align} 
Combining Eqs. (\ref{Th1})--(\ref{Th3}), we arrive at 
the result in the lemma. $\Box$

\begin{flushleft}
{\it
Completion of proof of Theorem \ref{QP+}
}
\end{flushleft} 
Theorem \ref{QP+} immediately follows from Lemmas \ref{Smallness} and \ref{Thermo}. $\Box$

\subsection{Proof of Theorem \ref{SP-}}
Because  $\displaystyle 
U\sum_{\bj\in \Lambda} q_{\bj}^2\ge 0
$,
 we have, from  Eq. (\ref{PP2}), 
\begin{align}
 \frac{\la -\tilde{H}\ra_{\Lambda}}{|\Lambda|}
\le 
8t +
4V \la  q_{{\bf o}} q_{{\boldsymbol
 \delta_1}}\ra_{\Lambda}-\frac{\Delta}{2} \la q_{{\bf o}} \ra_{\Lambda}.
\end{align}
Because  $|\la  q_{{\bf o}} q_{{\boldsymbol \delta_1}}\ra| \le
 \la  q_{{\bf o}}^2\ra $,  $
|\la  q_{{\bf o}} \ra |
\le \la  q_{{\bf o}}^2\ra^{1/2}
$
and 
$
\la  q_{{\bf o}}^2\ra
\le \la  q_{{\bf o}}^2\ra ^{1/2}
$, we obtain 
\begin{align}
\frac{\la -\tilde{H}\ra_{\Lambda}}{|\Lambda|}
\le 
8 t +\Big(
4V+\frac{\Delta}{2} \Big)\la  q_{{\bf o}}^2 \ra_{\Lambda}^{1/2}. \label{SInq}
\end{align}
We can easily verify that 
$\displaystyle 
\la\Omega |\tilde{H} \Omega\ra =\Big(|S|-\frac{\Delta}{2}\Big) |\Lambda|
$.
Thus, by the Peierls--Bogoliubov inequality, we have
$\displaystyle 
\ln \Tr_{\mathfrak{H}}[e^{-\beta \tilde{H}}]
\ge  \beta \Big(\frac{\Delta}{2}-|S|\Big) |\Lambda|
$. By combining this result with  Eqs.  (\ref{Th3}) and (\ref{SInq}),
we arrive at 
\begin{align}
\la q_{{\bf o}}^2 \ra_{\Lambda}^{1/2}
\ge
\frac{\frac{\Delta}{2}-|S|}{\frac{\Delta}{2}+4V} 
-\frac{8 t }{\frac{\Delta}{2}+4 V}
-\frac{2 \ln 4} {\beta(\frac{\Delta}{2}+4 V)}.
\end{align}
 Thus, for any $\vepsilon>0$, we have $
  \Big|
  1-
  \la q_{{\bf o}}^2\ra_{\Lambda}
  \Big|<\vepsilon$, provided that $\beta$ and $t^{-1}$ are
  sufficiently large.
The application of  Lemma \ref{Smallness} concludes the assertion. $\Box$

\appendix
\section{ Dyson--Lieb--Simon inequality}\label{DLS}
\setcounter{equation}{0}
Let $\mathfrak{X}_L$ and $\mathfrak{X}_R$ be complex Hilbert spaces.
For simplicity, we suppose that $\dim \mathfrak{X}_L=\dim \mathfrak{X}_R<\infty$.
Let $\vartheta$ be an antiunitary transformation from $\mathfrak{X}_L$
onto $\mathfrak{X}_R$.
Let $A, B_1, \dots, B_n\in \mathfrak{B}(\mathfrak{X}_L)$. Assume that
$A$
is self-adjoint.
Here we address the  self-adjoint operator defined by 
\begin{align}
H=A\otimes \one +\one \otimes \vartheta A\vartheta^{-1}
-\sum_{j=1}^n \Big(B_j\otimes \vartheta B_j \vartheta^{-1}+
B_j^*\otimes \vartheta B_j^* \vartheta^{-1}
\Big).
\end{align} 
As usual, thermal expectation associated with $H$ is given by 
\begin{align}
\la \bullet \ra=\Tr_{\mathfrak{X}_L\otimes \mathfrak{X}_R}[\bullet
\ e^{-H}]\Big/\Tr_{\mathfrak{X}_L\otimes \mathfrak{X}_R}[ e^{-H}].
\end{align} 

The following theorem is a   generalized version of the  DLS  inequality:
\begin{Thm}\label{DLSCl}
Let $C, D\in \mathfrak{B}(\mathfrak{X}_L)$. We have
\begin{itemize}
\item[{\rm (i)}] $
\la C\otimes \vartheta C\vartheta^{-1}\ra\ge 0,
$
\item[{\rm (ii)}]
$\displaystyle 
\Big|
\big\la C\otimes \vartheta D\vartheta^{-1}\big\ra
\Big|
\le
 \big\la C\otimes \vartheta C\vartheta^{-1}\big\ra
 \big\la D\otimes \vartheta D\vartheta^{-1}\big\ra.
$
\end{itemize} 
\end{Thm} 
\begin{rem}
{\rm 
\begin{itemize}
\item[(i)] 
In the original DLS inequality \cite{DLS},  all of the matrix elements of $A$ and
 $B_j$
 are assumed to be real. However, noted   in 
Refs. \cite{LiebNach, Miyao1,T. Miyao5},
 we can weaken this assumption.
\item[(ii)] As noted in Ref. \cite{Miyao6}, item (i) can be regarded as 
 a non-commutative  version of the  Griffiths inequality.  $\diamondsuit$
\end{itemize} 

}
\end{rem}

 Theorem \ref{DLSCl} can be proven by applying a series of lemmas.
The basic idea of our proof comes from  Ref. \cite{Miyao1}.

We begin with the following observation:
\begin{lemm}\label{Base1}
For each $A, B\in \mathfrak{B}(\mathfrak{X}_L)$, we have
\begin{align}
\Tr_{\mathfrak{X}_L\otimes \mathfrak{X}_R}
\big[
A\otimes \vartheta B \vartheta^{-1}
\big]
=\Tr_{\mathfrak{X}_L}[A]
( \Tr_{\mathfrak{X}_L}[B])^*.
\end{align} 
In particular, we have
\begin{align}
\Tr_{\mathfrak{X}_L\otimes \mathfrak{X}_R}
\big[
A\otimes \vartheta A \vartheta^{-1}
\big]
=\big|
\Tr_{\mathfrak{X}_L}[A]
\big|^2\ge 0.
\end{align} 
\end{lemm} 
{\it Proof.}
It suffices to prove that 
$
\Tr_{\mathfrak{X}_R}
\big[\vartheta B\vartheta^{-1}\big] =(\Tr_{\mathfrak{X}_L}[B])^*
$.
Let $\{e_i\}_{i}$  be a complete orthonormal system in  $\mathfrak{X}_R$.
Then $\{\vartheta^{-1} e_i\}$ is a complete orthonormal system in 
$\mathfrak{X}_L$ as well. We have
\begin{align}
\Tr_{\mathfrak{X}_R}\big[\vartheta B\vartheta^{-1}\big]
&=\sum_{i} \la e_i|\vartheta B\vartheta^{-1} e_i\ra
=\sum_{i} \la \vartheta \vartheta^{-1}e_i|\vartheta B\vartheta^{-1}
 e_i\ra
=\sum_{i} \big(\la \vartheta^{-1 }e_i| B\vartheta^{-1} e_i\ra\big)^*\no
&=(\Tr_{\mathfrak{X}_L}[B])^*.
\end{align} 
This completes the proof. $\Box$
\medskip\\

Let $\mathfrak{C}_0$ be a convex cone defined by 
\begin{align}
\mathfrak{C}_0=\mathrm{Coni}\Big\{
A\otimes \vartheta A\vartheta^{-1}\ \Big|\, 
A\in \mathfrak{B}(\mathfrak{X}_L)
\Big\},
\end{align} 
where $\mathrm{Coni}(S)$ is the conical hull of $S$.
Let $\mathfrak{C}$ be the closure of $\mathfrak{C}_0$ under the operator
norm topology.  A linear operator $X$ on $\mathfrak{X}_L\otimes \mathfrak{X}_R$
 is called {\it reflection positive} if $X$ belongs to $\mathfrak{C}$.
If $X$ is reflection positive, then we write $X\succeq 0$.

By Lemma \ref{Base1}, we have the following:
\begin{lemm}\label{RPPPP}
If $X\succeq 0$, then $\Tr_{\mathfrak{X}_L\otimes \mathfrak{X}_R}
[X] \ge 0
$.
\end{lemm} 

The following lemma is often useful:

\begin{lemm}
If $X\succeq 0, Y\succeq 0$, then we have the following
\begin{itemize}
\item[{\rm (i)}] $XY\succeq 0$;
\item[{\rm (ii)}] $a X+bY\succeq 0$ for all $a, b\ge 0$.
\end{itemize} 
\end{lemm}

\begin{Prop}\label{Base2}
We have $e^{-\beta H} \succeq 0$ for all $\beta \ge 0$.
\end{Prop} 
{\it Proof.}
Let 
\begin{align}
H_0=A\otimes \one +\one \otimes \vartheta A\vartheta^{-1},\ \ \ 
V=\sum_{j=1}^n\big(
B_j\otimes \vartheta B_j\vartheta^{-1}
+B_j^* \otimes \vartheta B_j^*\vartheta^{-1}
\big).
\end{align} 
First,  observe that 
$
e^{-\beta H_0}=e^{-\beta A} \otimes \vartheta e^{-\beta A}
 \vartheta^{-1}\succeq 0
$
for all $\beta \in \BbbR$.
Conversely, because  $V\succeq 0$, we have
\begin{align}
e^{\beta V}=\sum_{n=0}^{\infty} \underbrace{\frac{\beta^n}{n!}}_{\ge
 0}\underbrace{V^n}_{\succeq 0} \succeq 0
\end{align} 
for all $\beta \ge 0$. By the Trotter--Kato product formula, we obtain
\begin{align}
e^{-\beta H} =\lim_{n\to \infty}\Big(
\underbrace{e^{-\beta H_0/n}}_{\succeq 0} 
\underbrace{e^{\beta V/n}}_{\succeq 0}
\Big)^n \succeq 0
\end{align} 
for all $\beta \ge 0$. $\Box$

\begin{Prop}\label{Base3}
Assume that  $X\succeq 0$. For each $C, D\in
 \mathfrak{B}(\mathfrak{X}_L)$, we have
\begin{align}
\Big|
\Tr_{\mathfrak{X}_L\otimes \mathfrak{X}_R}
\big[
C\otimes \vartheta D\vartheta^{-1} X
\big]
\Big|^2
\le
\Tr_{\mathfrak{X}_L\otimes \mathfrak{X}_R}
\big[
C\otimes \vartheta C\vartheta^{-1} X
\big]
\Tr_{\mathfrak{X}_L\otimes \mathfrak{X}_R}
\big[
D\otimes \vartheta D \vartheta^{-1} X
\big].
\end{align} 
\end{Prop}
{\it Proof.}
For simplicity, we assume that $X\in \mathfrak{C}_0$. Thus, we can write
$X$
 as $X=\sum_{j=1}^N E_j\otimes \vartheta E_j\vartheta^{-1}$. By Lemma
 \ref{Base1}, we have
\begin{align}
\Tr_{\mathfrak{X}_L\otimes \mathfrak{X}_R}
\big[
C\otimes \vartheta D\vartheta^{-1} X
\big]
=&\sum_{j=1}^N \Tr_{\mathfrak{X}_L\otimes \mathfrak{X}_R}
\big[
CE_j \otimes \vartheta DE_j\vartheta^{-1} 
\big]\no
=& \sum_{j=1}^N \Tr_{\mathfrak{X}_L}[CE_j] \Big(
\Tr_{\mathfrak{X}_L} [DE_j]
\Big)^*.
\end{align} 
Hence, by the Schwartz inequality, we obtain
\begin{align}
\Big|\Tr_{\mathfrak{X}_L\otimes \mathfrak{X}_R}
\big[
C\otimes \vartheta D\vartheta^{-1} X
\big]\Big|^2
\le&
 \sum_{j=1}^N\Big|
\Tr_{\mathfrak{X}_L}[CE_j]
\Big|^2\sum_{j=1}^N\Big|
\Tr_{\mathfrak{X}_L}[DE_j]
\Big|^2\no
\le &
\Tr_{\mathfrak{X}_L\otimes \mathfrak{X}_R}
\big[
C\otimes \vartheta C\vartheta^{-1} X
\big]
\Tr_{\mathfrak{X}_L\otimes \mathfrak{X}_R}
\big[
D\otimes \vartheta D \vartheta^{-1} X
\big].
\end{align} 
 This finishes the proof. $\Box$

\begin{flushleft}
{\it Proof of Theorem \ref{DLSCl}}
\end{flushleft} 
By Proposition \ref{Base2}, we have
$
C\otimes \vartheta C\vartheta^{-1} e^{-\beta H} \succeq 0
$ for all $\beta \ge 0$. Therefore, by Lemma \ref{RPPPP}, we conclude
item (i).

Item (ii) immediately follows from Propositions \ref{Base2} and
\ref{Base3}. $\Box$

\end{document}